\documentclass[traditabstract]{aa} % for the abstract without structuration 
\usepackage{longtable,lscape}
\usepackage{graphicx}
\usepackage{txfonts}
\usepackage{natbib}
\bibpunct{(}{)}{;}{a}{}{,} % to follow the A&A style
\begin{document}
   \title{Insights into thermonuclear supernovae from the incomplete Si-burning process}

   \author{E. Bravo
%           \inst{\ref{inst1}}
           }

   \institute{Dept. F\'\i sica i Enginyeria Nuclear, Univ. Polit\`ecnica de
              Catalunya, Carrer Pere Serra 1-15, 08173 Sant Cugat del Vall\`es, Spain\\   
              \email{eduardo.bravo@upc.edu} 
%	      \label{inst1}
	      }

   \date{Received ; accepted }

% \abstract{}{}{}{}{} 
% 5 {} token are mandatory
 
  \abstract
{
Type Ia supernova (SNIa) explosions synthesize a few tenths to several tenths of a solar mass,
whose
composition is the result of incomplete silicon burning that reaches peak temperatures of 4 GK to 5
GK.
The elemental abundances are sensitive to the physical conditions in
the explosion, making their measurement a promising clue to uncovering the properties of the
progenitor star and
of the explosion itself. Using a parameterized description of the thermodynamic history of
matter undergoing incomplete silicon burning, we computed the final composition for a
range of parameters wide enough to encompass current models of SNIa. Then, we searched for
combinations of
elemental abundances that trace the parameters values and are potentially measurable. 
For this purpose, we divide the present study into two epochs of SNIa, namely the {\sl optical}
epoch, from a few weeks to several months after the explosion, and the {\sl X-ray} epoch, which
refers to the time period in which the supernova remnant is young, starting one or two hundred
years age and ending a thousand years after the event. During the optical epoch, the only SNIa
property that can
be extracted from the detection of incomplete silicon burning elements is the neutron
excess of the progenitor white dwarf at thermal runaway, which can be determined through
measuring the ratio of the abundance of manganese to that of titanium, chromium, or vanadium.
Conversely, in the
X-ray epoch, any abundance ratio built using a couple of elements from titanium, vanadium,
chromium, or manganese may constrain the initial neutron excess. Furthermore, measuring
the ratio of the abundances of vanadium to manganese in the X-ray might shed light on the
timescale of the thermonuclear explosion.
}


   \maketitle
%
%________________________________________________________________

\section{Introduction}

Type Ia supernovae (SNIa) are the brightest phenomena powered by nuclear reactions in the universe,
releasing $\sim1.5\times10^{51}$~erg that are one-third invested in overcoming the binding energy
of the progenitor white dwarf (WD), and the rest in kinetic energy of the debris. The nuclear
energy release takes roughly one second, implying a mean power of
$\sim1.5\times10^{44}$~W. These figures make SNIa prime targets for studies of nucleosynthesis in
extreme conditions of density and temperature. 

While there is currently little doubt that the star exploding as a SNIa is a binary WD
\cite[e.g.][]{nug11}, other details of the progenitor system are uncertain; to cite a few:
the nature
of the companion star \cite[either double degenerate, DD, or single degenerate,
SD;][]{pat07,mao08,blo09,sim09,li11,blo12,fol12,mao12,sch12}, the
mass-accretion history \citep{cho12,dis12,sha12},
the geometry of the flame at thermal runaway \citep{hoe02b,zin09}, the mass of the WD at
ignition \cite[either a sub-Chandrasekhar WD, a
Chandrasekhar-mass WD, or a super-Chandrasekhar one,][]{maz07,van10,yua10}, or the properties
of the environment within a parsec of the WD. In the SD
Chandrasekhar-mass WD scenario, the structure is supported by the pressure of degenerate
electrons, which can be described by a few parameters. As a consequence, only a handful
of the progenitor system properties have an impact on the thermonuclear explosion, namely the
central density of the WD, the initial geometry of the flame, and the chemical composition. 
The thermal content may also have an impact \citep{cal12}. Current knowledge of stellar
evolution implies that the progenitor WD has to be made up mainly of carbon and oxygen, in
proportions that depend on its main-sequence mass and metallicity. The metallicity has
further influence on the WD structure through setting the electron mole number,
$Y_{\mbox{e}}$, which controls the number of electrons available to support the WD. It is to be
expected that nature makes SNIa from a range of all the above parameters.

The workings of the thermonuclear bomb that powers a SNIa are not precisely known. Within the
SD
Chandrasekhar-mass WD scenario, the favored model involves an initially subsonic flame
(deflagration) followed by a detonation. The release of thermonuclear energy in either 
burning-propagation mode is what controls the time schedule of the explosion. Thus, there is an
intimate
relationship between the final energy and mechanical structure of the supernova ejecta, on the one
hand, and its
chemical composition and profile, on the other hand. Furthermore, the chemical structure of the
ejecta reflects the
thermodynamic conditions in the explosion. Thus, there would be possible in principle to make an
inverse analysis of the ejecta composition to uncover the details of the supernova
explosion. 

From the observational point of view, obtaining the chemical composition of the ejecta is not an
easy task. Optical spectra depend in a complex way on the physical conditions in the ejecta days to
weeks after the supernova explosion, which in turn are controlled by the energy deposition
from radioactive nuclei, mainly \element[][56]{Ni} and its decay product \element[][56]{Co}.
However, recently there have been advances that allow performing detailed tomographies of a few
well-studied SNIa 
\citep{ste05b,alt07,maz08,tan10b}. It is to be expected
that the set of SNIa for which a detailed chemical composition of the ejecta is available 
will
increase steadily in the next several years. 

An alternative way to gain knowledge about the chemical
structure of the ejecta is through studying young supernova remnants \citep{bad10}. 
\cite{bad08a} studied the relationship between X-ray spectral features due to Mn and Cr in
young
SNIa remnants and the metallicity of the WD progenitor, which
allowed them to conclude that the metallicity of the progenitor of Tycho's supernova is
supersolar. The same technique has been since applied to other suspected SNIa remnants to
determine their progenitors' metallicity.

The picture of SNIa ejecta drawn from observations has converged to a layered chemical structure,
although with
exceptions and peculiarities \cite[e.g.][]{fol12b}. In the center, there is 
a volume filled with relatively neutron-rich iron-group elements \citep{hoe04}, thought to arise
from
matter heated to more than $\sim5.5$ billion degrees which achieved nuclear statistical
equilibrium (NSE) and
thereafter was neutronized efficiently by electron captures. This neutronization clears the center
of
the main radioactive product of SNIa, \element[][56]{Ni}. Above the central core, there is a volume
made of iron group elements that achieved NSE at a density low enough to
experience a negligible amount of neutronization. This volume is mainly composed
of \element[][56]{Ni}.
Outwards from this
region, there is a volume made of the products of incomplete silicon burning (hereafter,
Si-b), covering a wide range in atomic number, from silicon to iron. Finally,
close to the surface of the ejecta, there is a tiny region made of unburned fuel and
material that has been processed  no further than oxygen burning. The details of the
layered structure vary from object to object, and not all the observations point to the same
degree of chemical stratification. Some objects seem to have experienced a thorough mixing of the
two outer layers, unburned and Si-b matter \citep{ste05}, although there is evidence that
iron-group
nuclei and the products of Si-b have clearly different physical histories \citep{rak06}.

Although the masses of the different layers of SNIa ejecta are thought to
vary from object to object, it can be said that NSE matter should amount from 0.2 to
0.8~M$_{\sun}$,
whereas intermediate-mass elements, mainly made as a result of Si-b, cover more or less the same
range of masses \citep{maz07}. In principle, the study of both regimes therefore has the capability
of
providing interesting clues to the nature of the explosion. However, whereas the abundances of NSE
matter are influenced by the amount of electron captures achieved in the first phases of the
explosion, the composition resulting from Si-b only depends on the initial composition of the
supernova progenitor and on the dynamics of the explosion. 

The purpose of the present work is to elucidate the dependence of the chemical composition arising
from Si-b on the progenitor properties (metallicity and chemical composition of the exploding WD)
and on the explosion dynamics (expansion timescale and combustion wave velocity).
We note that, although incomplete silicon burning has been studied in depth in quite a few works
\cite[e.g.][]{bod68,woo73,hix96,hix99,mey96,mey98}, none of them have looked at the final
chemical composition for tracers of the parameters governing this nucleosynthetic process.
We use a model of the thermodynamic evolution of matter undergoing Si-b, which we explain and
justify in
the next section. Thereafter, we focus on abundance ratios of Si-b products, which allows us to
avoid the problem posed by the dependence of the absolute abundances on the total energy of the
supernova. We split the results into two groups of abundance ratios: on the one hand, those that
might create a detectable signature in the optical display of the supernova, encompassing
between a few days and a few months after the explosion, which we term the {\sl optical phase} or
the 
{\sl optical epoch}; and on
the other hand, those that would apply to supernova remnants several hundred years after the
event, which we term the {\sl X-ray phase} or the {\sl X-ray epoch}, because abundance
measurements 
in this phase are usually obtained from X-ray spectra. We also refer to the first second of
the
explosion 
as the {\sl nucleosynthetic epoch}, since it is then when the nuclear fluxes are able to modify the
final 
chemical composition, apart from radioactive disintegrations. 

\section{Numerical method}

We performed the nucleosynthetic calculations described in this paper by solving the nuclear
kinetics equations with the code CRANK described in \cite{bra12}. For completeness, we summarize
here its main features. CRANK integrates the temporal evolution of the abundances in a
nuclear network for given thermal and structural (density) time profile,
and for initial composition. The inputs to CRANK are the nuclear data and the thermodynamic
trajectories, as a function of time. The thermodynamic trajectories, in turn, can be the result of
a hydrodynamic calculation of an explosive event, in which case CRANK provides the final chemical
profile of the ejecta after computing the nucleosynthesis of each mass shell independently,
without accounting for chemical diffusion. Alternatively, CRANK can compute the
nucleosynthesis in a single homogeneous region using an arbitrary thermodynamic trajectory. In the
present work, we use CRANK in the latter way.

The nuclear
evolutionary equations \cite[see e.g.][]{rau02}
follow the time evolution of the molar fraction of each nucleus until the temperature falls below
$10^8$~K, after which time the chemical composition is no longer substantially modified, with the
exception of radioactive disintegrations. 
The integration of the nuclear network follows the implicit iterative method of
Wagoner with adaptive time steps. The iterative procedure ends when the relative variation in the
molar abundances is smaller than $10^{-6}$, taking only those nuclei with molar
abundance $Y>10^{-14}$~mol~g$^{-1}$ into account. The number of iterations is limited to 7, and if
this
limit is reached a new iteration procedure with a reduced time step is started from the abundances
at the last successful integration. In case the relative variation in the abundance
of any nuclei with $Y>10^{-14}$~mol~g$^{-1}$ is greater than a prescribed tolerance,
the iteration procedure is restarted from the last successful integration, but with a smaller time
step. The sum
of mass fractions is checked upon completion of each iteration procedure. If
that sum differs from one more than $10^{-8}$, the iteration procedure needs to be restarted with a
smaller time step. 

The network is adaptive, in the sense that its size (nuclei and reactions linking them) varies
during the computation in order to improve its performance.  
A nucleus enters the calculation only if either it has an appreciable
abundance ($Y>10^{-24}$~mol~g$^{-1}$) or it
can be
reached from any of the abundant nuclei by any one of the reactions included in the network. Light
particles (neutrons, protons, and alphas) are the exception to the rule, because they are always
included in the network. 
A reaction rate is included in the network only if the predicted change it induces on a molar
abundance in
the next time step is larger than a threshold, fixed at $10^{-20}$~mol~g$^{-1}$
\cite[see][for a similar method]{rau02}.

The nuclear network consists of a maximum of 722 nuclei, from free nucleons up to $^{101}$In,
linked by three fusion reactions: $3\alpha$, $^{12}$C+$^{12}$C,
and $^{16}$O+$^{16}$O, electron and positron captures, $\beta^-$ and $\beta^+$ decays, and 12
strong interactions per each nucleus with $Z\ge6$, involving neutrons, protons, alphas, and
$\gamma$.
We take the necessary nuclear data from \cite{sal69,ito79,ffn82,mar00,cyb10}.

\section{Modeling the physical conditions in incomplete silicon burning in SNIa}

Explosive nucleosynthesis studies can be divided into two broad categories, namely those that
follow the
thermodynamic trajectories computed from hydrodynamic models of the exploding object
\cite[e.g.][]{thi86,mae10}, and those that adopt a simplified parameterized description of the
time dependence of the relevant thermodynamic variables \citep{woo73,mey98,hix99}. The former
ones are more adequate for determining the chemical composition of the ejecta of a particular
supernova
(or whatever exploding object) model. The second ones are suitable for studies aiming at
elucidating nucleosynthetic properties expected to be shared by a general class of explosion
models. Here, we chose this second approach because we are interested in just one nucleosynthetic
process among all that are at work in SNIa.

As is customary in parameterized studies of high-temperature explosive nucleosynthesis (op cit),
we model the temporal evolution of the mass zone undergoing Si-b as a fast heating followed by
adiabatic expansion and cooling. The temperature as a function of time is given by

\begin{equation}\label{eqT}
 T_9 = \left\{ \begin{array}
                {r@{\quad:\quad}l}
		T_{9,0}+\left(T_{9,\mbox{peak}}-T_{9,0}\right)t/\tau_{\mbox{rise}} &
0\le t\le \tau_{\mbox{rise}} \\
		T_{9,\mbox{peak}}\exp\left[-\left(t-\tau_{\mbox{rise}}\right)/\tau\right] &
\tau_{\mbox{rise}} < t \\
               \end{array} \right.
\,,
\end{equation}

\noindent where $t$ is time, $T_9$ is the temperature in units of $10^9$~K, and the rest of the
quantities are model parameters, namely $\tau_{\mbox{rise}}$ is the time to reach the maximal
temperature,
$T_{9,\mbox{peak}}$, starting from an initial value, $T_{9,0}$, and $\tau$ is the expansion
timescale. In some studies, the expansion timescale has been taken as the free fall timescale,
$\tau_{\mbox{ff}}=446/\sqrt{\rho}$; however, we prefer to keep it as a free parameter in the
present work. The initial temperature, $T_{9,0}$ is irrelevant as long as it is low enough compared
to the peak temperature. We fixed its value at $T_{9,0}=1$ in all the numerical
experiments. 

The density is given by

\begin{equation}\label{eqro}
 \rho_7 = \left\{ \begin{array}
                {r@{\quad:\quad}l}
		\rho_{7,0}+\left(\rho_{7,\mbox{peak}}-\rho_{7,0}\right)t/\tau_{\mbox{rise}} &
0\le t\le \tau_{\mbox{rise}} \\
		\rho_{7,\mbox{peak}}\exp\left[-3\left(t-\tau_{\mbox{rise}}\right)/\tau\right] &
\tau_{\mbox{rise}} < t \\
               \end{array} \right.
\,,
\end{equation}

\noindent where $\rho_7$ is the density in units of $10^7$~g~cm$^{-3}$, $\rho_{7,\mbox{peak}}$
is the maximal density, and $\rho_{7,0}$ is the initial density, whose value is not influential in
the present results and is fixed in the calculations at $\rho_{7,0}=0.3\rho_{7,\mbox{peak}}$.
The factor three in the exponential in Eq.~\ref{eqro} accounts for the $\rho\propto T^3$
dependence, appropriate to describing the isentropic evolution of a radiation-dominated gas
\citep{hix99}. 

Although adiabatic expansion is a common assumption, matter heated to temperatures high enough to
achieve partial or total nuclear statistical equilibrium, $T\ga4\times10^9$~K, may go off
the isentropic thermodynamic path because of the continuous adjustment of the chemical composition
as temperature declines, which releases nuclear energy. 
The degree to which Eqs.~\ref{eqT} and \ref{eqro} provide a faithful representation of the
nucleosynthesis in SNIa layers undergoing Si-b can be tested, for any given SNIa model, by
computing
the chemical composition from these equations with parameters fitted to the thermodynamic
trajectories provided by the hydrodynamic code, then comparing it to the chemical composition
obtained with the original thermodynamic trajectories. 

\begin{figure}[tb]
\centering
   \includegraphics[width=9 cm]{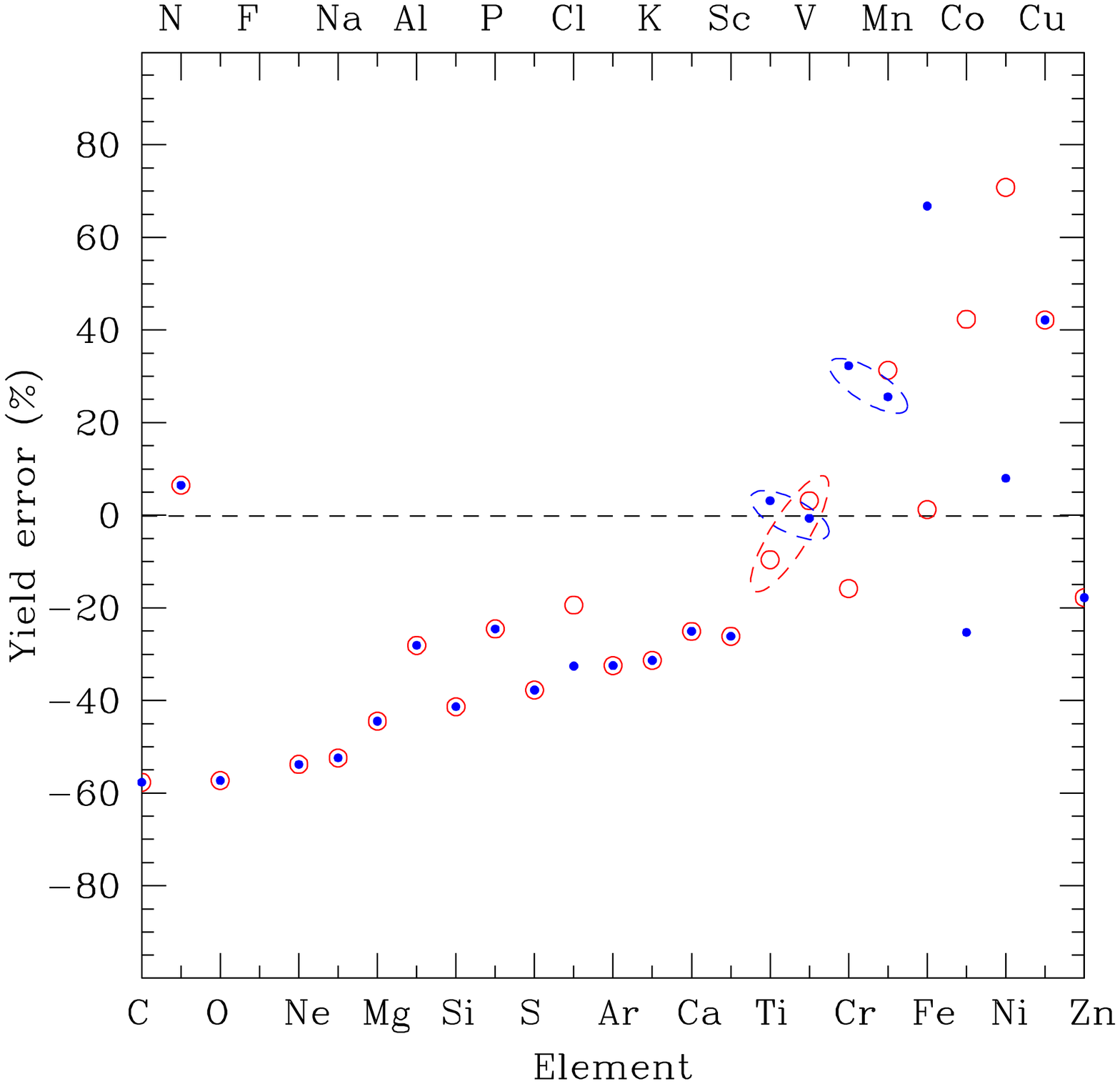}
\caption{(Color online)
Error in the elemental yields obtained with Eqs.~\ref{eqT} and \ref{eqro} relative to those
 for the same element obtained with the thermodynamic trajectories provided by a hydrodynamic code
for a SNIa model (see text for further details). Large open (red) circles give the errors belonging
to a time of one day after the explosion, when the abundances of many iron-group elements are still
dominated by radioactive isotopes. Small filled (blue) circles give the errors belonging to a time
one hundred years after the explosion. Points belonging to some interesting couples of elements
are highlighted by a small ellipse around them. For this comparison, we have only
considered the layers
with peak temperatures between 4.3 and 5.2 GK.
}
\label{fig1}
\end{figure}

Figure~\ref{fig1} compares the yields obtained for the same elements with the two thermodynamic
time evolutions described in the previous paragraph, for the SNIa model DDTc described in
\cite{bad05c} and \cite{bra12} and used hereafter as a reference model. As can be deduced from the
figure, the errors in the abundances of elements from silicon to niquel related to the use of
Eqs.~\ref{eqT} and \ref{eqro} range from 
-40\% to +70\%. Most important, taking ratios of the 
yields of different elements, the errors can be made very tiny. For instance, the errors of the
yields of the elements belonging to the first quasi-statistical equilibrium (QSE) group, i.e.
those between silicon and scandium, are all very similar, implying that the ratio of the yields
obtained
with Eqs.~\ref{eqT} and \ref{eqro} are a very good representation of the ratios belonging to
the full thermodynamic trajectories provided by the supernova model. The errors of the elements
belonging to the second QSE
group, i.e. those between titanium and zinc, are much more heterogeneous. However,
couples of elements can be identified in Fig.~\ref{fig1} whose yields are affected by a very
similar error, hence for whom the yield ratios are predicted well by using Eqs.~\ref{eqT} and
\ref{eqro}. An example is the couple formed by titanium and vanadium, at both one day and one
hundred years after the explosion. Another example is the
couple formed by chromium and manganese at one hundred years after the explosion.

%We did also try fitting to the thermodynamic histories of
%supernova layers more complex analytical functions, searching for a smaller error, however
%we obtained no improvement with respect to
%the single exponential dependence on time representative of adiabatic expansion. 

We also tried fitting the thermodynamic histories of supernova layers with more complicated
algebraic functions, but we obtained no
improvements over the single exponential
dependence on the adiabatic expansion timescale.

\subsection{Parameters ranges}\label{parrange}

With the temporal evolution of temperature and density given by Eqs.~\ref{eqT} and \ref{eqro}, the
full set of parameters that controls the final chemical composition of mass layers undergoing
incomplete Si-b 
are the
peak temperature $T_{9,\mbox{peak}}$, the peak density $\rho_{7,\mbox{peak}}$, the expansion
timescale $\tau$, the temperature rise timescale $\tau_{\mbox{rise}}$, the initial mass fraction
of \element[][12]{C},
$X(\element[][12]{C})$, and the initial mass fraction of \element[][22]{Ne},
$X(\element[][22]{Ne})$.
The last two quantities provide a convenient parameterization of the chemical composition of the
WD at runaway. In this work, we simply assume that the only species initially present in the WD are
\element[][12]{C},
\element[][16]{O}, and \element[][22]{Ne}. The initial mass fraction of oxygen is thus
\begin{equation}
 X(\element[][16]{O}) = 1 - X(\element[][12]{C}) -
X(\element[][22]{Ne})\,,
\end{equation}
\noindent and the initial neutron excess is 
\begin{equation}\label{eqeta}
 \eta = \frac{X(\element[][22]{Ne})}{11}\,.
\end{equation}
\noindent In turn, the initial neutron excess is related to the metallicity of the supernova
progenitor, $Z$, through $\eta\approx0.101Z$ \citep{tim03}, which leads to, $Z\approx
X(\element[][22]{Ne})/1.11$\footnote{It should be recalled that the relationship between the
neutron excess of the progenitor WD at runaway and the metallicity of the progenitor star at
zero-age main-sequence 
can be modified owing to electron captures during the hundreds of years of the
carbon-simmering phase that precedes thermal runaway \citep{pir08b,bad08a}.\label{foot1}}. 

In the following, we discuss the relevant ranges of the parameters to explore.

\subsubsection{Peak temperature}

In our reference model, DDTc, as well as in most SNIa models
\citep[e.g.][]{woo73,hix99,mae10},
incomplete Si-b is achieved at densities slightly in excess of $10^7$~g~cm$^{-3}$
\citep[see Fig. 1 in][]{bra12}. At these densities, to achieve the QSE of the iron group and
of the silicon group, the peak temperature has to exceed $\sim4\times10^9$~K. On the
other hand, to avoid reaching full NSE the temperature should remain well below
$\sim5.5\times10^9$~K
\citep{woo73}. As we demonstrate later, the most promising targets of the present study
are the lightest elements of the iron-group, from titanium to manganese. As can be
seen in Table~2 of \cite{bra12}, the bulk of these elements originates in SNIa in layers
attaining
temperatures in the range from $4.2\times10^9$~K to $5.2\times10^9$~K, with the exception of
manganese, which has a sizeable contribution from layers reaching NSE. For the purposes of the
present study, we define the range of peak temperatures able to provide incomplete Si-b such as
that between $4.3 - 5.2\times10^9$~K.

\begin{figure}[tb]
\centering
   \includegraphics[width=9 cm]{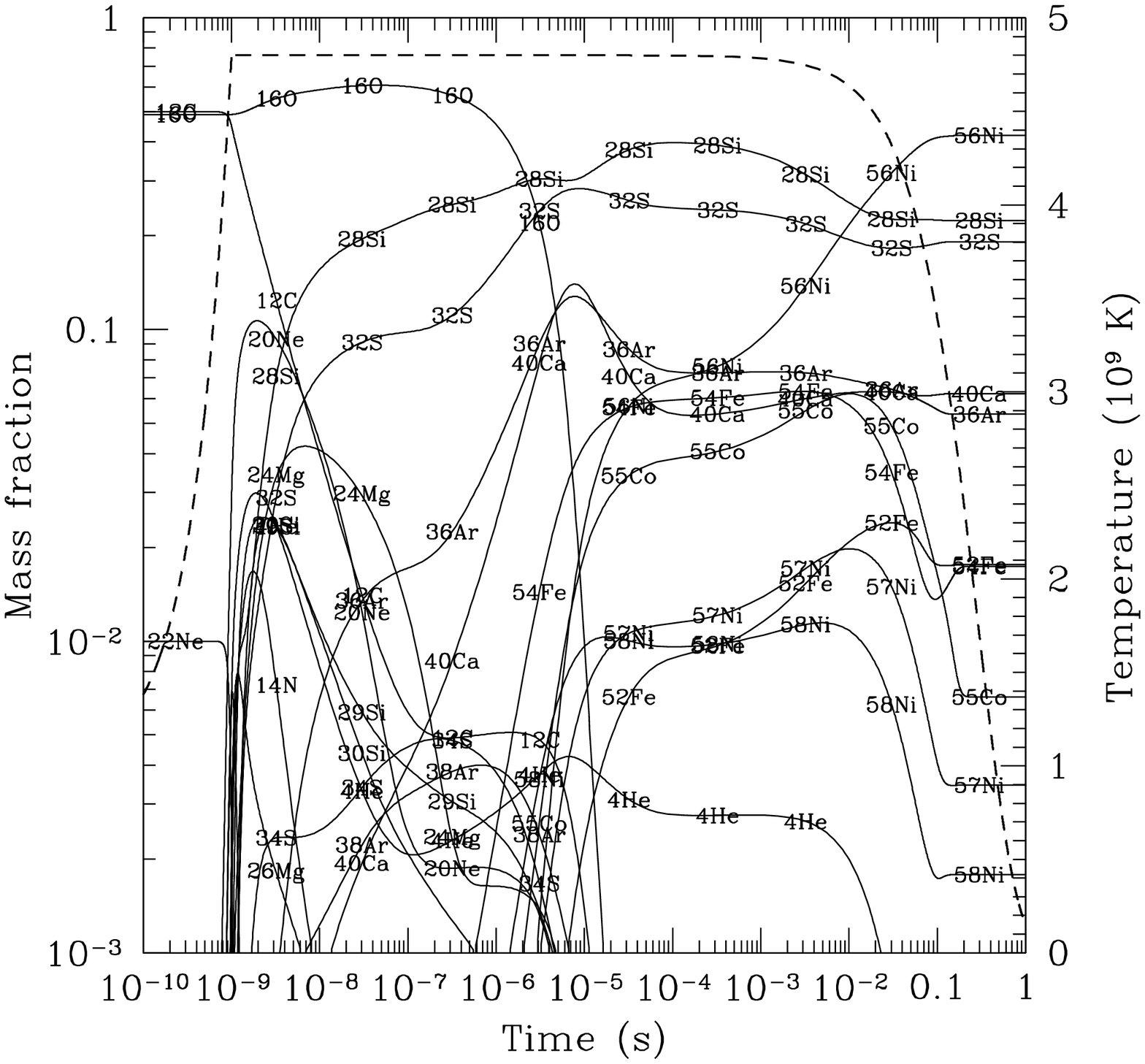}
\caption{Evolution of the chemical composition for a reference set of parameters, namely
$T_{\mbox{peak}}=4.8\,10^9$~K, $\tau_{\mbox{rise}}=10^{-9}$~s, $\tau=0.29$~s,
$X(\element[][12]{C})=0.5$, $X(\element[][22]{Ne})=0.01$. The dashed line is the temperature in
$10^{9}$~K.}
\label{fig2}
\end{figure}

Figure~\ref{fig2} shows the evolution of the nuclear species computed with
$T_{9,\mbox{peak}}=4.8$, $\rho_{7,\mbox{peak}}=2.2$, $\tau=0.29$~s,
$\tau_{\mbox{rise}}=10^{-9}$~s, $X(\element[][12]{C})=0.5$, and $X(\element[][22]{Ne})=0.01$. For
this
set of parameters, and in spite of the high temperature attained, there is no appreciable change
in the chemical composition before $t=\tau_{\mbox{rise}}$. Carbon is exhausted in a time of
a few tens of ns and oxygen burns at less than 10~$\mu$s, whereas the isotopes belonging to the
silicon QSE-group achieve stable proportions after $\sim0.1$~ms. It can be seen that the
species of the iron QSE-group steadily increase in abundance after $\sim1$~ms and until the
expansion cools matter below $\sim4\times10^9$~K.
After this point, two different trends in the
isotopes belonging to the iron QSE-group can be discerned. On the one hand, \element[][56]{Ni} is
the only
species that continues growing in abundance. On the other hand, the mass fractions of the rest of
the most abundant components of the iron QSE-group, rich in neutrons, decrease owing to
recombination with excess free protons and alphas, adapting their abundances to equilibrium at the
decreasing temperatures, while they remain above $T\gtrsim3\times10^9$~K \citep{hix99}. This
recombination causes
a drop in the abundance of \element[][4]{He} after $\sim0.01$~s, which is clearly visible in the
plot.

It is notable that the final mass fractions of \element[][55]{Co} (\element[][55]{Mn} after a few
years of radioactive
disintegration) and \element[][52]{Fe} (grandparent of \element[][52]{Cr}) are among the largest
from the iron QSE-group.
But for \element[][54]{Fe} and \element[][56]{Ni}, they are the most abundant species
with $\eta\neq0$ and with $\eta=0$, respectively. 

\subsubsection{Distribution function of peak temperatures.}

\begin{figure*}[tb]
\centering
   \includegraphics[width=9 cm]{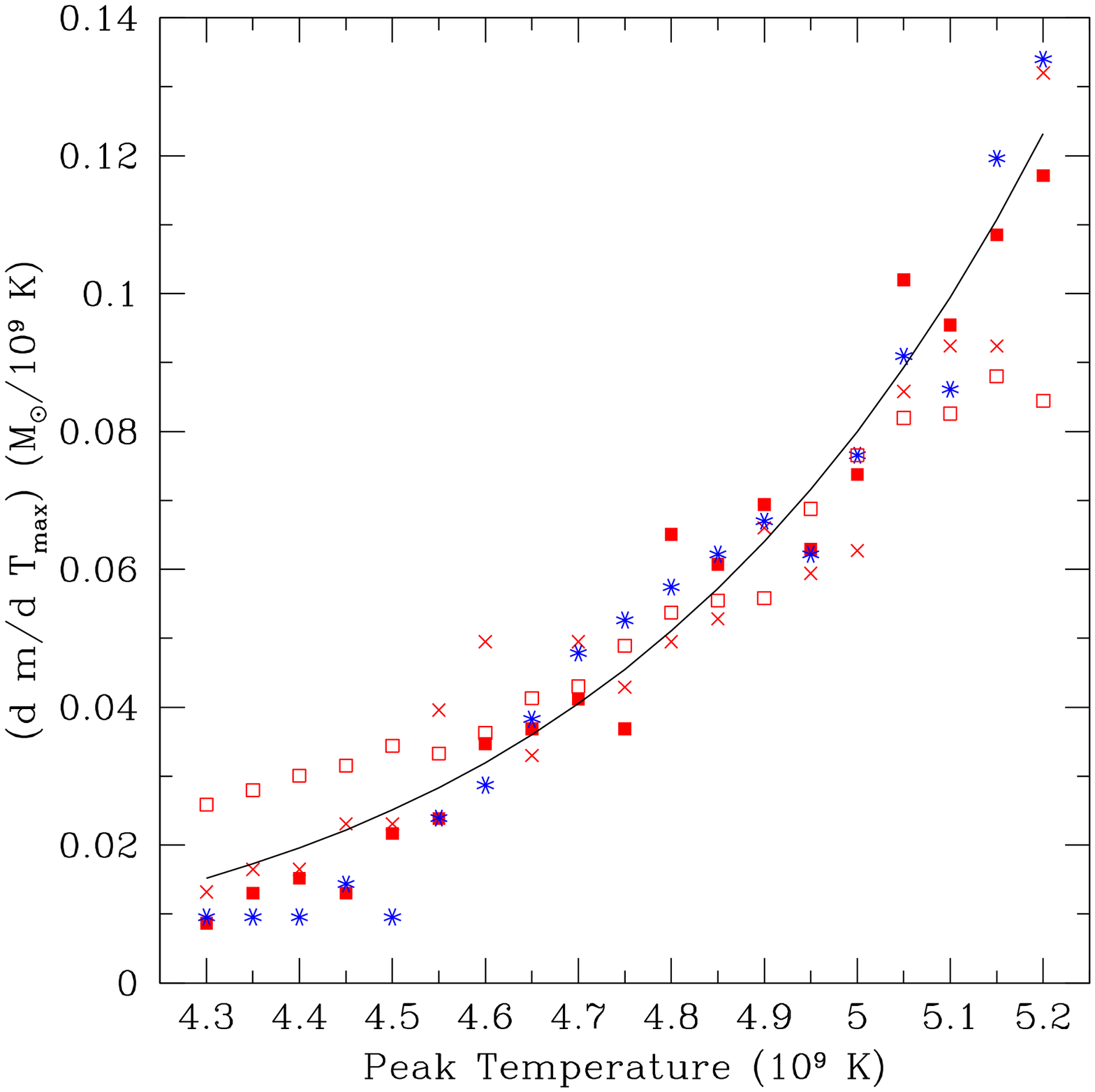} \includegraphics[width=9 cm]{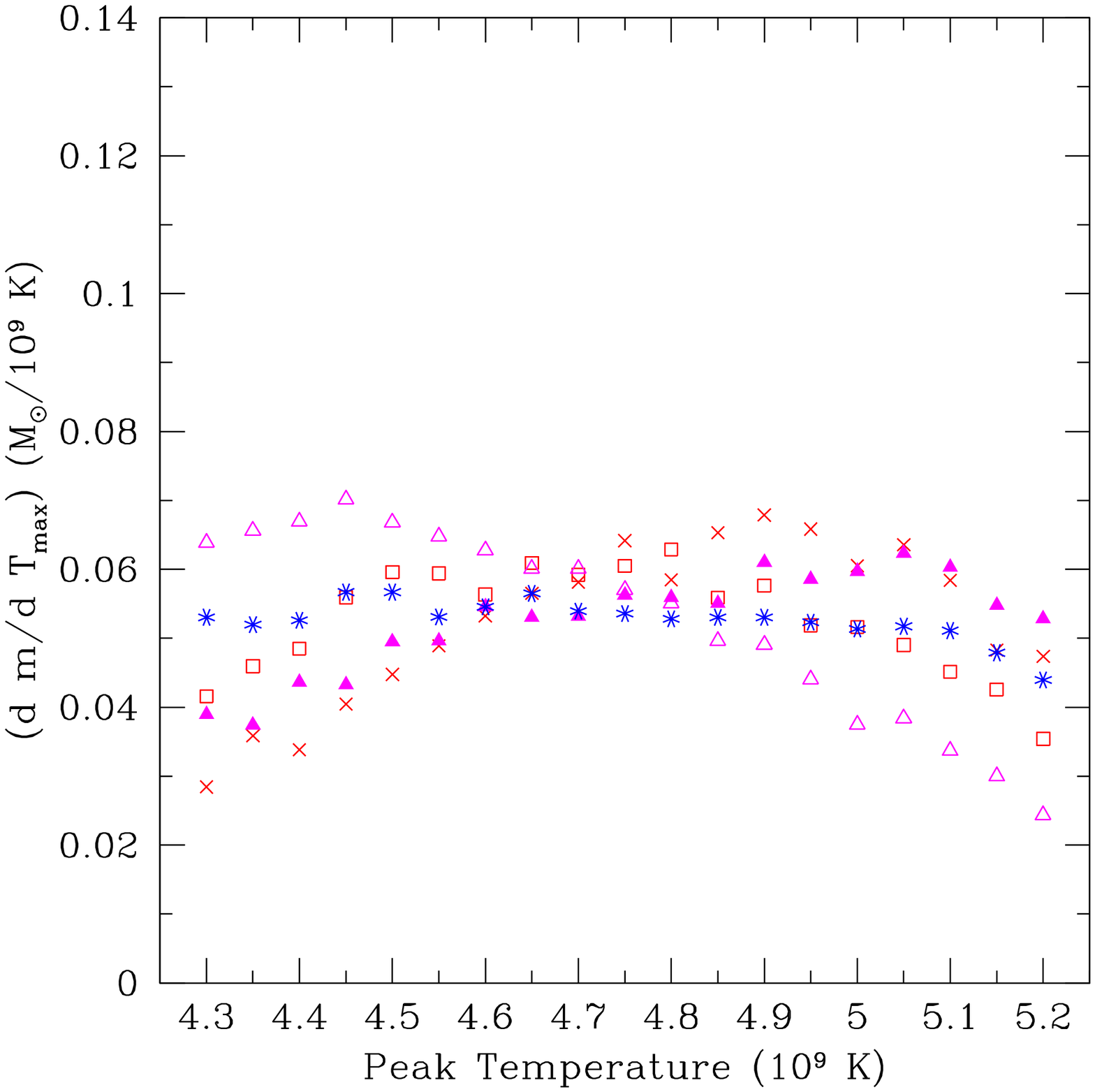}
\caption{(Color online) Distribution of maximal temperatures in the incomplete Si-burning region
obtained from a
suite of SNIa models. 
{\bf Left:} One-dimensional delayed detonation models DDTc \citep[red crosses,][]{bad05c}
and DDTf (red solid squares), a sub-Chandrasekhar model (blue stars) from
\citet[][fourth row in their Table 2]{bra11}, and three-dimensional delayed detonation
model TURB7 from \citet[][red empty squares,]{bra09b}. The solid line is a fit to the data given
by,
$\mbox{d}m/\mbox{d}T_{\mbox{peak}} \propto T_{\mbox{peak}}^{11}$. 
{\bf Right: } Three-dimensional delayed detonation models
DDT3DA (red empty squares) and DDT3DB (red crosses) from \citet{bra08}, gravitationally
confined detonation model GCD1 (magenta solid triangles), pulsating reverse
detonation model PRD18 (magenta empty triangles), both from \citet{bra09b}, and sub-Chandrasekhar
model C (blue stars) from \citet{gar99}. The distribution of maximal temperatures in these models
is approximately constant, independent of the peak temperature, unlike the models shown
in the left panel.
}
\label{fig3}
\end{figure*}

In any model of SNIa, matter attains a wide range of peak temperatures. As the
burning front moves through layers of decreasing density, its peak temperature drops. 
Matter experiencing Si-b is not subject to just a single peak temperature but to a sequence of
them. Figure~\ref{fig3} shows the distribution of peak temperatures in a suite
of SNIa models of different flavors. The distribution function of peak temperatures in these
models can be described by either an approximately power-law dependence on
$T_{9,\mbox{peak}}$, which we fit with $\mbox{d}m/\mbox{d}T_{\mbox{peak}}
\propto T_{\mbox{peak}}^{11}$, or as independent of $T_{9,\mbox{peak}}$. Aside from the models
shown in these figures, model O-DDT in
\cite{mae10} follows a dependence on $T_{9,\mbox{peak}}$ that is similar to the proposed
power law, at least qualitatively, whereas those of models W7 and C-DEF (op cit) are 
described better as independent of $T_{9,\mbox{peak}}$. 

In the following, we use these two representative distribution functions of the peak
temperatures to estimate the mass of the different elements created by given combinations of the
parameters of Si-b evolution, and we analyze the impact of the peak-temperature
distribution function on the results. If this impact were
sizeable, observations might provide insight into the 
distribution function of peak temperatures in SNIa. If, on the other hand, its impact is
negligible, the
results will be independent of the assumed distribution function. 

\subsubsection{Peak density}

Peak density in Si-b depends on the peak temperature, both quantities being
correlated with entropy. \cite{mey98} introduced a quantity proportional to the entropy
per nucleon in photons,
\begin{equation}\label{eqphi}
 \phi\equiv3.4\times10^{-3}T_9^3/\rho_7\,.
\end{equation}
\noindent Given the thermodynamic history described by Eqs.~\ref{eqT} and \ref{eqro}, $\phi$
remains
constant. 

For model DDTc, in the region undergoing incomplete Si-b the peak density,
$\rho_{7,\mbox{peak}}$ can be approximated as a function of the peak temperature by the
relationship

\begin{equation}\label{eqropeak}
 \rho_{\mbox{7,peak}} = \exp\left(T_{\mbox{9,peak}}-4\right)\,.
\label{eq1}
\end{equation}

\noindent Using this equation, $\phi$ ranges from 0.15 to 0.2, which are much lower values than
investigated
in \cite{mey98}. In SNIa, the peak temperature and density are the result of the thermodynamic
conditions in matter that is either shocked in a detonation wave (delayed-detonation models) or
burnt
subsonically
(deflagration models). \cite{kho88} provides the peak temperature and density for Chapman-Jouguet
deflagration and detonation waves, which differ from Eq.~\ref{eqropeak} by at most 22\% in the
temperature and density ranges of Si-b. 

\begin{figure}[tb]
\centering
   \includegraphics[width=9 cm]{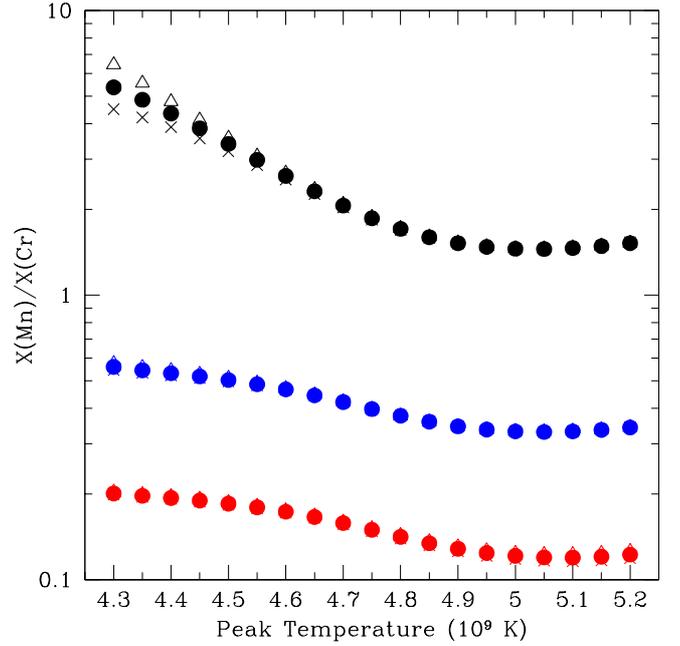}
\caption{(Color online) Final mass fraction ratio Mn/Cr, at a time of 100~yrs after the explosion,
as a
function of peak temperature in incomplete
Si-b for different maximal densities (point type) and initial neutron excesses (color
online). The maximal densities are given by either
Eq.~\ref{eqropeak} (filled circles), or by a factor of two higher (triangles), or lower (crosses)
than Eq.~\ref{eqropeak}. The
initial neutron excesses are $\eta=2.27\,10^{-4}$ (red), $9.09\,10^{-4}$ (blue), and
$6.82\,10^{-3}$ (black), increasing from bottom to top.}
\label{fig4}
\end{figure}

To reduce the number of free parameters, we use Eq.~\ref{eqropeak} in the calculations
reported in the present work. It is therefore necessary to evaluate the degree of uncertainty
derived from this relationship. This is done in Fig.~\ref{fig4}, where we represent the ratio from
manganese to chromium at the X-ray epoch, for peak densities given by either Eq.~\ref{eqropeak},  a
factor of two higher or a factor of two lower. It is clearly seen that the precise value of the
peak
density is irrelevant to the mass ratio of these elements. The points belonging to the
different peak densities used are nearly indistinguishable in the plot, except at the highest
neutron excess and the lowest temperature. These last points belong to the highest
entropies studied in the present work. It is interesting to note that \cite{mey98} explored even
highest entropies and neutron excesses. 

\subsubsection{Rise time}\label{secdefdet}

Most, if not all, nucleosynthetic studies of the explosive incomplete Si-b process compute the
nuclear flows starting from the peak temperature and the initial fuel composition. \cite{woo73}
argue that this is justified when the rise time, i.e. the time taken to dump
the energy needed to raise its temperature to its peak value into the mass zone, is much shorter
than the
expansion timescale, $\tau_{\mbox{rise}}\ll\tau$. In SNIa models, the energy dumped into a zone
until it reaches its peak temperature can either be due to a
compressional heating related to the passage of the shock front at the leading edge of a detonation
wave or it is a combination of thermal diffusion and energy released by nuclear reactions in a
deflagration wave. The rise time associated with each one of these burning waves is very different.

\begin{figure}[tb]
\centering
   \includegraphics[width=9 cm]{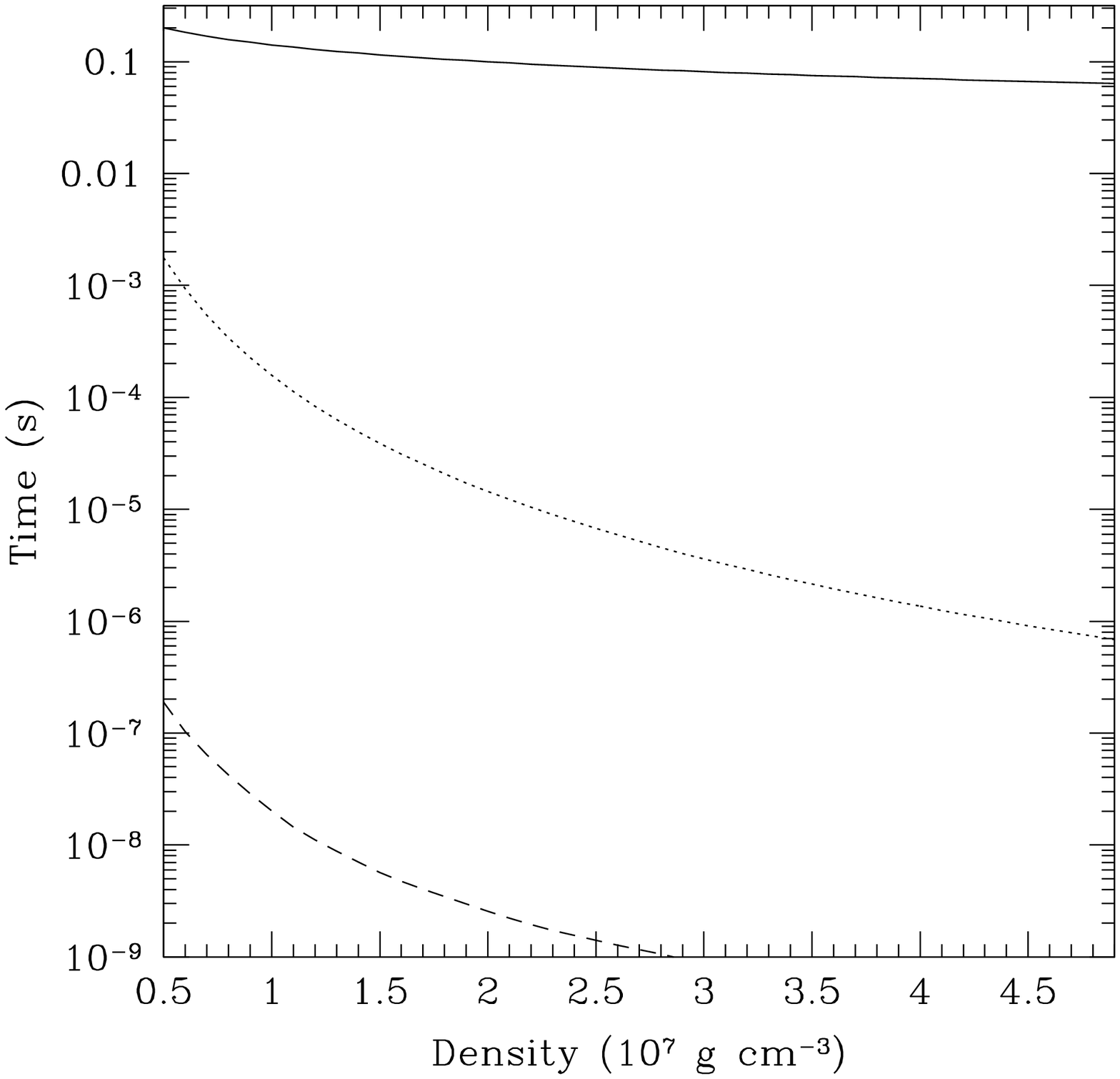}
\caption{
Comparison of relevant timescales as a function of the initial density. Solid line: expansion
timescale, computed as the free-fall or hydrodynamic timescale, $446/\sqrt{\rho}$~s;
dotted line: rise time in a deflagration wave, computed as the nuclear timescale at the critical
temperature at which the nuclear timescale matches the heat diffusion timescale; dashed line: rise
time in a detonation wave, computed as the nuclear timescale at the temperature of shocked matter.
}
\label{fig5}
\end{figure}

Figure~\ref{fig5} shows a comparison of both rise times as a function of density, together with the
hydrodynamic timescale. 
The thermal structure of a laminar flame
(deflagration) is determined by heat diffusion from hot ashes to cool fuel at low temperature and
by the release of heat by nuclear reactions at high temperature. The transition between both
regimes takes place at a critical temperature, $T_{\mbox{crit}}$, that depends on density. The
deflagrative timescale in Fig.~\ref{fig5} has
been estimated as the nuclear timescale at $T_{\mbox{crit}}$. For simplicity, both
$T_{\mbox{crit}}$ and the nuclear timescale were computed, accounting only for the dominant
reaction rate, i.e. the fusion of two $^{12}$C nuclei. 

Triggering of ignition in a supersonic
combustion wave (detonation) is due to an initial temperature jump in a length scale of a few mean
free paths, in a shock front. Later on, the heat
released by thermonuclear reactions determines the thermal profile. Thus, the detonation timescale
can be estimated as the nuclear timescale at the shock temperature, $T_{\mbox{shock}}$. For the
calculations shown in Fig.~\ref{fig5}, the shock temperature was calculated by solving the
Hugoniot jump conditions using the velocities of Chapman-Jouguet detonations given in \cite{kho88}.

The detonation rise time is the
shortest one by orders of magnitude, whereas the deflagration rise time lies between the detonation
rise time and the hydrodynamic timescale.
An alternative way of estimating the
timescale of a deflagration is as the ratio between the flame width and velocity. Using the flame
properties reported in \cite{tim92}, the deflagration timescale goes into the range $9\times10^{-3}
- 10^{-4}$~s for densities between $10^7$ and $5\times10^7$~g~cm$^{-3}$. Although these timescales
are about one order of magnitude longer than the ones in Fig.~\ref{fig5}, the qualitative
conclusions remain the same.

In view of the large difference between both rise times
and the hydrodynamic timescale, it is to be expected that the precise value of
$\tau_{\mbox{rise}}$ will be unimportant for the nucleosynthesis.
Notwithstanding this expectation, we have explored the nucleosynthesis resulting from different
rise times ranging from $10^{-9}$~s to $0.2$~s, in search for possible, although improbable,
tracers of the type of burning wave responsible for incomplete Si-b in SNIa.

\begin{figure*}[tb]
\centering
   \includegraphics[width=9 cm]{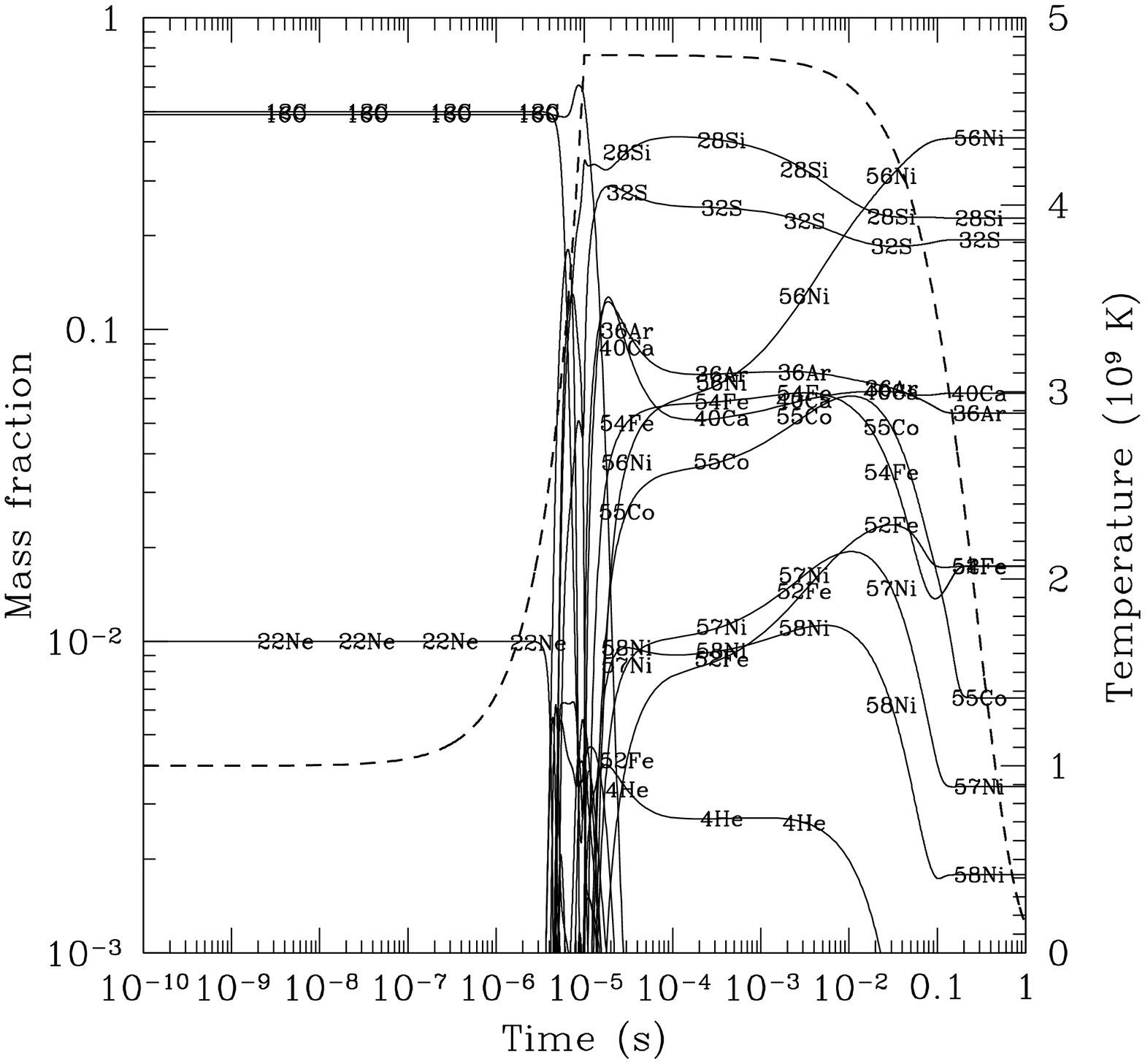} \includegraphics[width=9 cm]{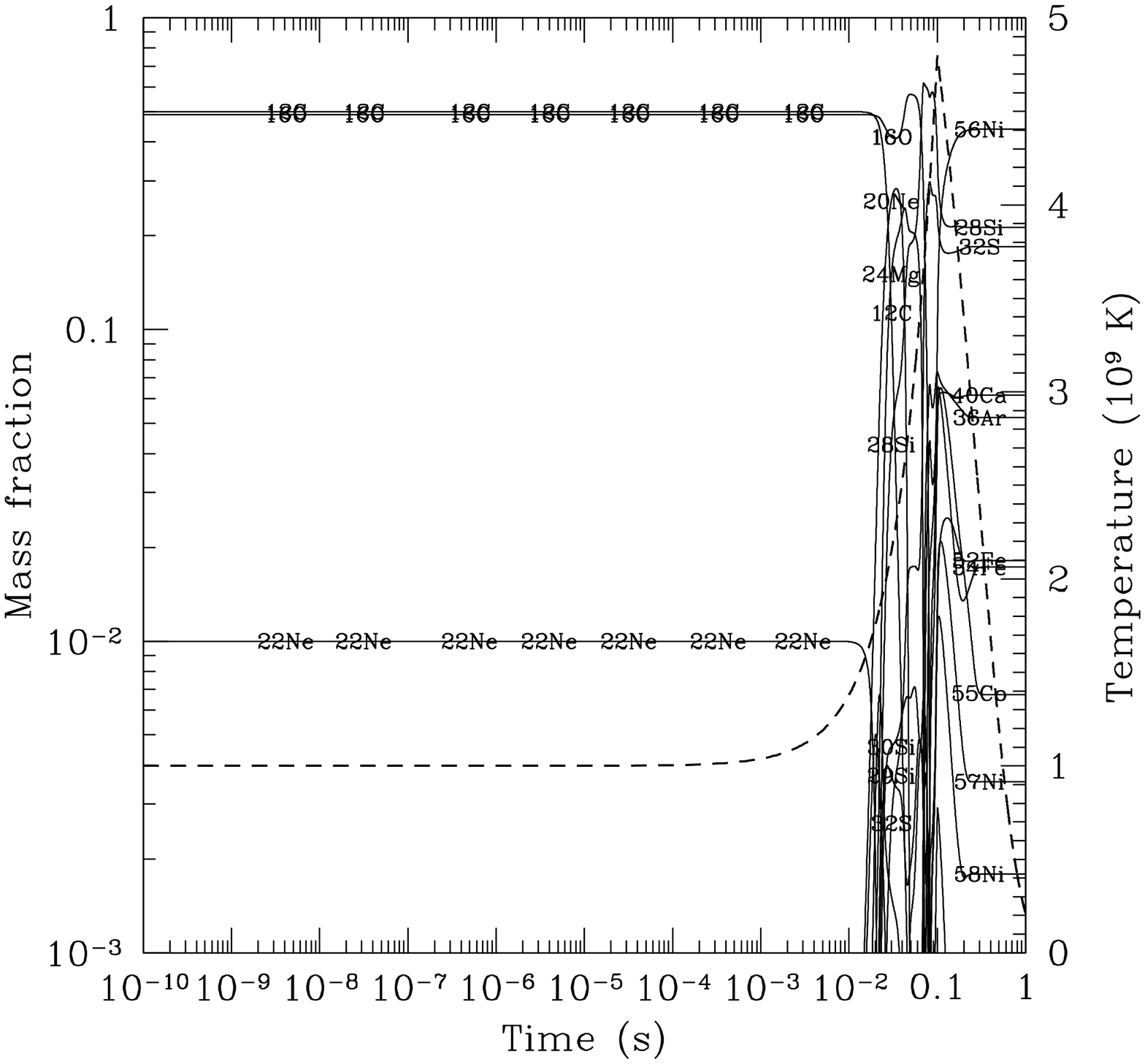}
\caption{Evolution of the chemical composition for a rise time to the maximal temperature
different from that in the reference set of parameters (Fig.~\ref{fig2}). {\bf Left:} Slower rise
to the peak temperature, $\tau_{\mbox{rise}}=10^{-5}$~s. {\bf Right:} Rise time comparable to the
expansion timescale, $\tau_{\mbox{rise}}=0.1$~s.}
\label{fig6}
\end{figure*}

Figure~\ref{fig6} shows the evolution of nuclear species for conditions similar to those in
Fig.~\ref{fig2} apart from the rise time. In the left hand panel of this figure, the rise time has
been set to $10^{-5}$~s to
illustrate the case of a deflagration wave, whereas Fig.~\ref{fig2} is more
representative of a detonation wave. In spite of some differences in the evolution of the chemical
abundances, as for instance in the extent of burning before $T_{\mbox{9,peak}}$, the
final mass fractions in both figures are identical, in line with the predictions
of \cite{woo73}. Moreover, the same conclusion applies to a calculation with a rise time, 
$\tau_{\mbox{rise}}=0.1$~s, comparable to the expansion timescale (right panel in Fig.~\ref{fig6}).

\subsubsection{Expansion timescale}

In SNIa, the expansion timescale is bounded to be longer than the WD sound-crossing time, $\tau
\gtrsim t_{\mbox{sound}} = R_{\mbox{WD}}/v_{\mbox{sound}} \approx 0.1 - 0.2$~s, which is similar to
the
free-fall timescale at the densities of interest to incomplete Si-b. In the present work, we 
explore expansion timescales in the range $0.1 - 0.9$~s. 

\begin{figure*}[tb]
\centering
   \includegraphics[width=9 cm]{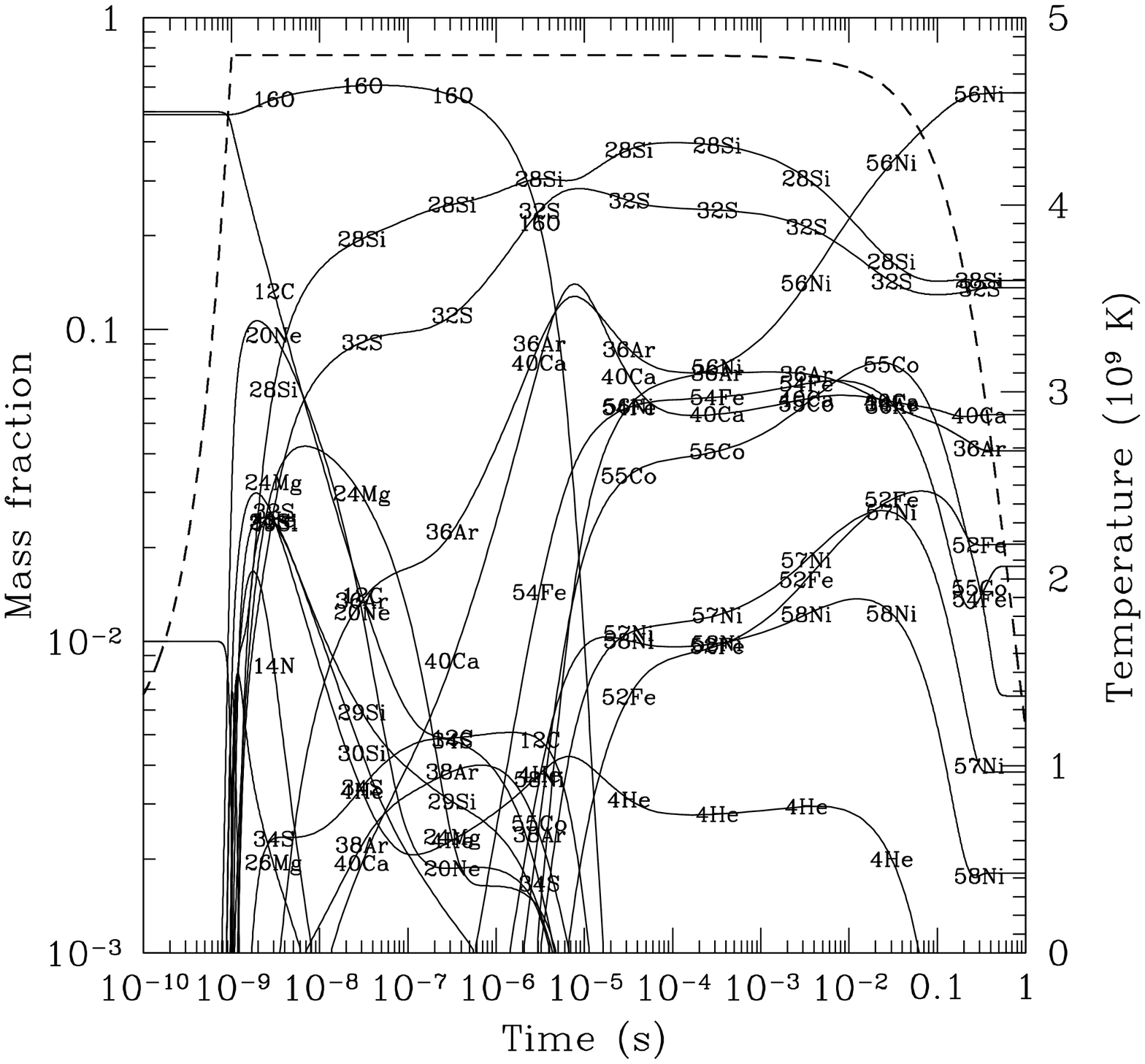}  \includegraphics[width=9 cm]{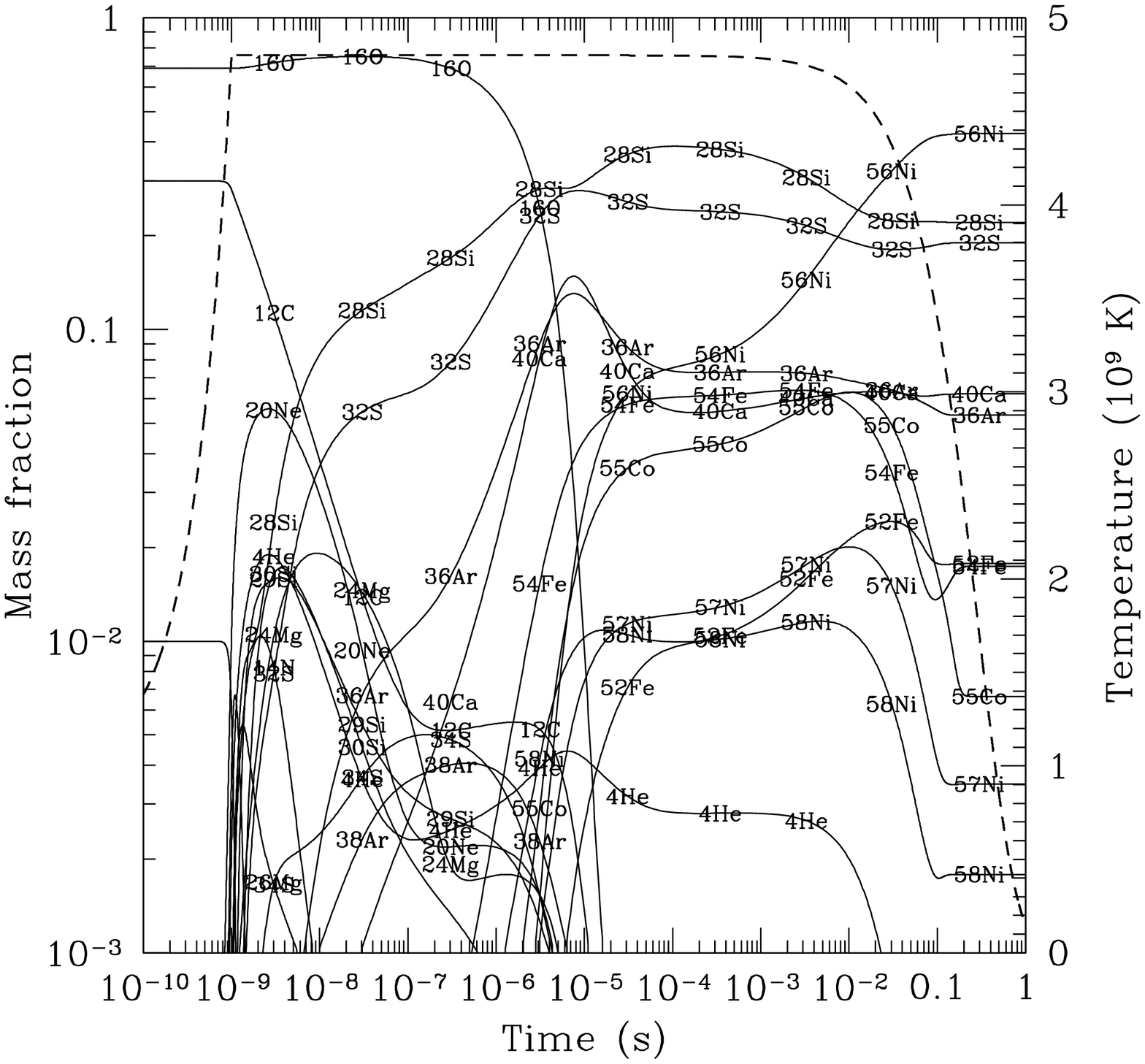}
\caption{Evolution of the chemical composition for different expansion timescales and initial
carbon abundances than in the reference set of parameters (Fig.~\ref{fig2}). {\bf Left:} Slower
expansion timescale, $\tau=0.72$~s. {\bf Right:} Smaller carbon abundance,
$X(\element[][12]{C})=0.3$.}
\label{fig7}
\end{figure*}

The left hand panel of Fig.~\ref{fig7} shows the evolution of the chemical composition for an
expansion timescale
$\tau=0.72$~s, i.e. slower than the one in Fig.~\ref{fig2}, $\tau=0.29$~s, while the rest of
parameters remain
unchanged. As could be expected, the differences only appear in the final phase of the evolution,
as can be seen in the temperature curve and in the leveling out of the abundances when all the
species leave QSE. Interestingly, a longer expansion timescale favors synthesis of a larger
quantity of \element[][56]{Ni} at the expense of \element[][28]{Si} and \element[][32]{S}, but the
rest of the species with
$X>0.001$ is hardly affected at all. 

\subsubsection{Initial carbon abundance}

\cite{dom01} accurately computed the evolution of stars in the range $1.5 - 7$~M$_{\sun}$ from main
sequence to the formation of a WD, starting from different metallicities. According to their
results, the carbon mass fraction in the WD lies in the range $0.37 - 0.55$.
We therefore explore a slightly wider range from $X(\element[][12]{C})=0.3$ to
$X(\element[][12]{C})=0.7$.

The right hand panel of Fig.~\ref{fig7} shows the evolution of the chemical composition for
initial abundances $X(\element[][12]{C})=0.3$ and $X(\element[][16]{O})=0.69$. Comparison with
Fig.~\ref{fig2}, in which $X(\element[][12]{C})=0.5$ and $X(\element[][16]{O})=0.49$, reveals that
the
initial carbon abundance has no influence on the final nucleosynthesis of the most abundant
elements. Once oxygen is exhausted, the evolution in both figures is identical. 

\subsubsection{Progenitor metallicity}

Type Ia supernovae are detected in host galaxies with a wide range of metallicities
\cite[e.g. $Z\approx0.1 - 2.5 Z_{\sun}$ in ][]{dan11}, with apparent preference for
hosts with $Z>Z_{\sun}$. The low end of the metallicity range might reaches
$Z\approx0.04
Z_{\sun}$ according to the preliminary results in \cite{rig11}. Here, we present results
of incomplete Si-b for initial \element[][22]{Ne} mass fractions between $2.5\times10^{-4}$ and
$7.5\times10^{-2}$, i.e. $Z\approx0.02 - 5 Z_{\sun}$ for a solar metallicity $Z_{\sun} \approx
0.012
- 0.014$ \citep{lod03,asp06}.

\begin{figure}[tb]
\centering
   \includegraphics[width=9 cm]{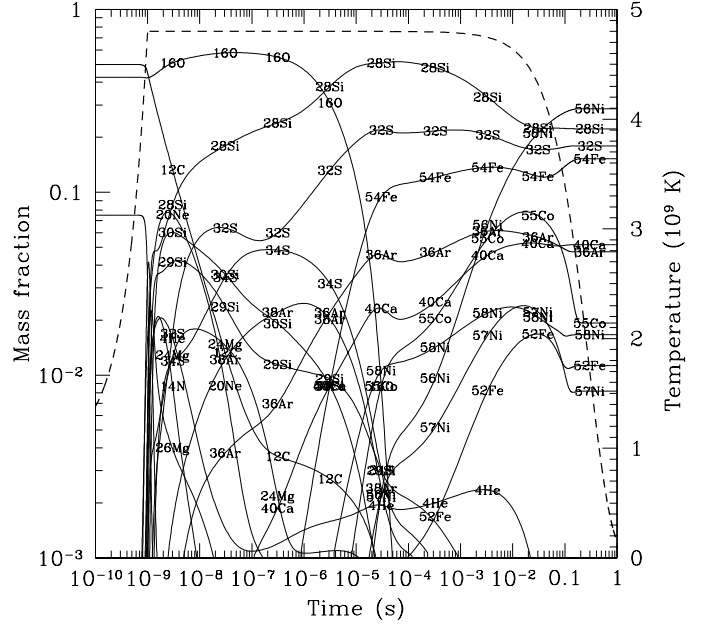}
\caption{Evolution of the chemical composition for an initial neutron excess higher than in the
reference set of parameters (Fig.~\ref{fig2}), $X(\element[][22]{Ne})=0.075$.}
\label{fig8}
\end{figure}

Figure~\ref{fig8} shows the evolution of the chemical composition starting from a neutron
excess $7.5$ times longer than in Fig.~\ref{fig2}. Comparing these two figures, the impact of
the initial neutron excess on the final abundances is evident. In Fig.~\ref{fig8}, the initial
amount of \element[][22]{Ne} disappears as soon as carbon is consumed. The role of
\element[][22]{Ne} as the main
neutron-excess holder is later played by $^{34}$S in oxygen burning, and by \element[][54]{Fe} and
\element[][55]{Co} in silicon burning and during freeze-out. 

\subsection{Summary of model parameters}

To summarize, four free parameters remain for which we try to find tracers from the
nucleosynthesis of incomplete Si-b, namely the temperature rise time, the expansion
timescale, the initial carbon abundance, and the progenitor
metallicity, $Z\approx X(\element[][22]{Ne})/1.11$. The initial values of temperature and density,
$T_{9,0}$ and $\rho_{7,0}$, are fixed, and the peak density is linked to the peak temperature
through Eq.~\ref{eq1}. Finally, the peak temperature is another parameter that is varied in
the different nucleosynthetic calculations, but such results are integrated using the
distribution function of peak temperatures. To be precise, in the following when we present the
results for a given set of parameters [$\tau_{\mbox{rise}}$, $\tau$, $X(\element[][12]{C})$, and
$Z$], the
mass fraction of each element is given by
%\begin{equation}\label{eqaverage}
% X = \int X(T_{\mbox{peak}}) \frac{\mbox{d}m}{\mbox{d}T_{\mbox{peak}}} \mbox{d}T_{\mbox{peak}}
%\propto \int X(T_{\mbox{peak}}) T_{\mbox{peak}}^n \mbox{d}T_{\mbox{peak}}\,,
%\end{equation}
\begin{eqnarray}\label{eqaverage}
 X & = \int X(T_{\mbox{peak}}) \left(\mbox{d}m/\mbox{d}T_{\mbox{peak}}\right)
\mbox{d}T_{\mbox{peak}} \nonumber\\
   & \propto \int X(T_{\mbox{peak}}) T_{\mbox{peak}}^n \mbox{d}T_{\mbox{peak}}\,,
\end{eqnarray}
\noindent with $\mbox{d}m/\mbox{d}T_{\mbox{peak}} \propto T_{\mbox{peak}}^n$, and either
$n=11$ or $n=0$ (Fig.~\ref{fig3}). 

\begin{table}[tb]
\caption{Sets of parameters of Si-b shown in Figs.~\ref{fig9}, \ref{fig11}, and
\ref{fig13}$-$\ref{fig15}\tablefootmark{a}.}\label{tab0}
\centering
\begin{tabular}{ll} 
\hline\hline             
Parameter & Values \\
\hline
$n$\tablefootmark{b} & 11, 0 \\
$\tau$ & (0.1, 0.2, 0.3, 0.4, 0.5, 0.6)/ln(2)~s \\
$X(\element[][22]{Ne})$ & $2.5\times10^{-4}$, $2.5\times10^{-3}$, 0.01, 0.025, 0.075 \\
$\tau_{\mbox{rise}}$ & $10^{-9}$, $10^{-7}$, $10^{-5}$, $10^{-3}$, 0.01, 0.05, 0.1, 0.2~s \\
$X(\element[][12]{C})$ & 0.3, 0.4, 0.5, 0.6, 0.7 \\
\hline
\end{tabular}
\tablefoot{
\tablefoottext{a}{All the combinations of the parameters values are represented in these figures
except when the contrary is stated in the corresponding captions.}
\tablefoottext{b}{Exponent of $T_{\mbox{peak}}$ in Eq.~\ref{eqaverage}. The integral in
Eq.~\ref{eqaverage} was approximated by summing the abundances obtained at
different temperatures, $T_9=4.3-5.2$ with interval 0.05 GK, where each point was weighted by
$T_{\mbox{peak}}^n$.}
}
\end{table}

\section{Tracers of the progenitor metallicity: X-ray epoch}

Since isotopic abundances cannot be measured in supernova ejecta, it is convenient to
know what isotopes are the main contributors to the abundances of elements made in incomplete
Si-b,
in the X-ray as well as in the optical epochs. This information is given in Table~\ref{tab1}.
When focusing on the X-ray phase, all the radioactivities that synthesize sizeable quantities of
elements from scandium to iron cease a long time before forming a supernova remnant, so
their elemental abundances remain constant at this epoch. 

The only pairs of elements whose mass
ratio in the X-ray epoch correlates strongly with progenitor metallicity are manganese
vs chromium,
manganese vs titanium, and vanadium vs titanium. We discuss them next, starting with the
ratio of manganese to chromium, which was originally proposed by \cite{bad08a} as a metallicity
tracer.

\begin{table*}[tb]
\caption{Parent species of Si-b elements at the optical and X-ray epochs.}\label{tab1}
\centering
\begin{tabular}{llcl} 
\hline\hline             
Element & Optical epoch\tablefootmark{a} & Variation\tablefootmark{b} & X-ray
epoch\tablefootmark{c} \\
\hline
%Si & $^{28}$Si & & $^{28}$Si \\
%P & $^{31}$P & & $^{31}$P \\
%S & $^{32}$S & & $^{32}$S \\
%Cl & $^{35}$Cl & & $^{35}$Cl \\
%Ar & $^{36}$Ar & & $^{36}$Ar \\
%K & $^{39}$K & & $^{39}$K \\
%Ca & $^{40}$Ca & & $^{40}$Ca \\
Sc & \element[][45]{Ti} (0.18 s) & $\sim$ & \element[][45]{Ti} (0.18 s) \\
Ti & \element[][48]{Cr} (21.6 h) $\rightarrow$ \element[][48]{V} (16.0 d) & $\uparrow$ &
\element[][48]{Cr} (21.6 h)
$\rightarrow$
\element[][48]{V} (16.0 d)\\
V & \element[][48]{Cr} (21.6 h) $\rightarrow$ \element[][48]{V} (16.0 d) & $\downarrow$ &
\element[][51]{Mn} (0.05 s)
$\rightarrow$
\element[][51]{Cr}    (27.7 d) \\
  & \element[][49]{Cr} (0.04 s) $\rightarrow$ \element[][49]{V} (330 d) & $\sim$ & \\
  & \element[][51]{Mn} (0.05 s) $\rightarrow$ \element[][51]{Cr} (27.7 d) & $\uparrow$ & \\
Cr & \element[][52]{Fe} (8.3 h) $\rightarrow$ \element[][52]{Mn} (5.6 d) & $\uparrow$ &
\element[][52]{Fe} (8.3 h) $\rightarrow$
\element[][52]{Mn} (5.6     d) \\
Mn & \element[][52]{Fe} (8.3 h) $\rightarrow$ \element[][52]{Mn} (5.6 d) & $\downarrow$ &
\element[][55]{Co} (17.5 h)
$\rightarrow$ \element[][55]{Fe} (2.7 y) \\
   & \element[][53]{Fe} (0.009 s) $\rightarrow$ \element[][53]{Mn} (3.7 Myr) & $\sim$ & \\
Fe & \element[][54]{Fe} & & \element[][56]{Ni} (6.1 d) $\rightarrow$ \element[][56]{Co} (77 d) \\
\hline
\end{tabular}
\tablefoot{
\tablefoottext{a}{Main contributor to the abundance of the element at the optical epoch (in
parenthesis, if it is an unstable isotope, its radioactive lifetime).}
\tablefoottext{b}{This column indicates if the contribution of a radioactive chain to the element
abundance in the optical phase  
increases with time ($\uparrow$), decreases ($\downarrow$), or remains about constant
($\sim$).}
\tablefoottext{c}{Main contributor to the abundance of the element at the X-ray epoch (in
parenthesis, if it is an unstable isotope, its radioactive lifetime).}
}
\end{table*}

\subsection{Ratio of manganese to chromium}\label{mn2cr}

\begin{figure}[tb]
\centering
   \includegraphics[width=9 cm]{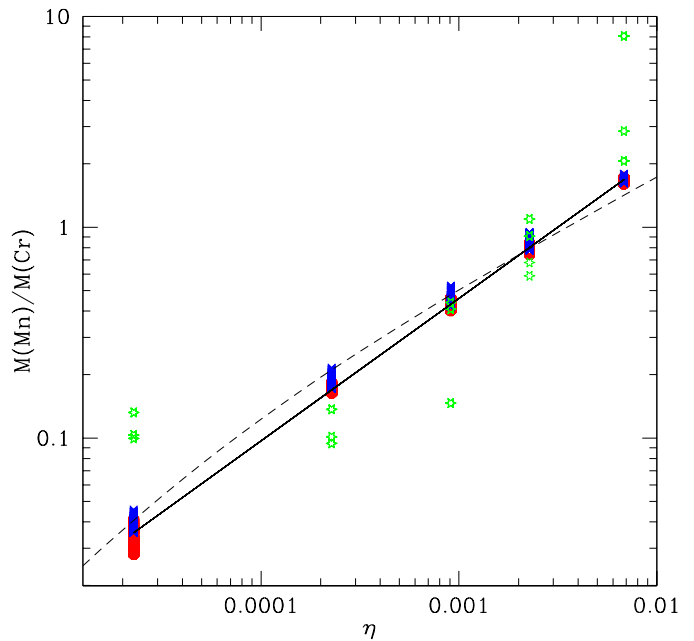}
\caption{(Color online) Averaged final mass fraction ratio  Mn/Cr as a function of the initial
neutron excess.
The average mass fractions are drawn either from the fit to the distribution of maximum
temperatures given in the left panel of Fig.~\ref{fig3} (red circles) or from a uniform
distribution function
representative of the models shown in the right panel of the same figure (blue crosses). In both
cases, the results are
shown for different $\tau_{\mbox{rise}}$ in the range $10^{-9}$-0.2~s,
$\tau$ in the range 0.10-0.90~s, and $X(\element[][12]{C})$ in the range 0.3-0.7. The
solid line is a fit given by, $M(\mbox{Mn})/M(\mbox{Cr})=49.1\eta^{0.676}$. The dashed line
shows the relationship expected from an analytic model, Eq.~\ref{eqanamncr}. 
Green stars give the mass fraction ratio  Mn/Cr for high-entropy conditions (see text for
details).
}
\label{fig9}
\end{figure}

Figure~\ref{fig9} shows the ratio of the abundances, either total masses or mass fractions, of
manganese to chromium (Mn/Cr) in the X-ray
epoch,
averaged by applying Eq.~\ref{eqaverage}, as a function of the initial neutron excess. The
insensitivity of the Mn/Cr
ratio to the temperature rise time and to the initial carbon abundance is
remarkable, as well as to the
distribution function of peak temperatures. At any given neutron excess, the final Mn/Cr ratio
varies
by less than a factor of two for the whole range of the rest of parameters, whereas the Mn/Cr ratio
as a
function of $\eta$ varies as much as a factor of 80. Furthermore, manganese and chromium are one of
the element pairs whose ratio is determined better with the simplified thermodynamic
evolution provided by Eqs.~\ref{eqT} and \ref{eqro} (see Fig.~\ref{fig1}). Thus, the
Mn/Cr ratio stands out as a robust tracer of the initial neutron excess at WD runaway. The Mn/Cr
ratio can be fit as a function of $\eta$ by a power law\footnote{
The statistical uncertainty
in the exponent of Eqs.~\ref{eqfitmncr} and \ref{eqfitmnti} is $\pm0.001$, that of
Eq.~\ref{eqfitvti} is $\pm0.003$, and that of Eqs.~\ref{eq22}, \ref{eq23}, and \ref{eq24} is
$\pm0.01$.}
(Fig.~\ref{fig9}),
\begin{equation}\label{eqfitmncr}
  M(\mbox{Mn})/M(\mbox{Cr})=49.1\eta^{0.676}\,.
\end{equation}
\noindent 
The results presented here are complementary to the work of \cite{bad08a}, where the
Mn/Cr ratio is computed for a series of SNIa models. We find a similar
power-law exponent as the one reported in that work, 0.676 vs 0.65, and the new power-law fit stays
within a factor of two
from the one in \cite{bad08a}. 

The robustness of the Mn/Cr ratio as a metallicity tracer is enhanced by its relative
insensitivity to the peak temperature (Fig.~\ref{fig4}). As a result, the Mn/Cr ratio is 
insensitive to the degree of mixing the nucleosynthetic yields for the supernova layers
undergoing incomplete Si-b. 

One can wonder if the Mn/Cr ratio can be used  as a metallicity tracer for core-collapse
supernovae. The answer is no, as long as layers experiencing Si-b in core collapse supernovae
evolve through a much higher isoentropic than in SNIa. We show in Fig.~\ref{fig9} the
results of the Mn/Cr mass ratio as a function of $\eta$ when the parameter $\phi$ defined in
Eq.~\ref{eqphi} 
takes the
value $\phi=6.8$, which is appropriate for core collapse supernovae \citep{mey96}, while $\tau$
varies in the range
$0.29 - 0.87$~s, and the range of $X(\element[][12]{C})$ is $0.3 - 0.7$. It can be deduced from
this figure that the tight relationship between the abundances of Mn and Cr and the metallicity is
a result of a low entropy evolution that cannot be extrapolated to core-collapse conditions. 

%\begin{figure}[tb]
%\centering
%   \includegraphics[width=9 cm]{fig13.eps}
%\caption{Final mass fraction ratio of Mn/Cr as a function of peak temperature during incomplete
%Si-burning for different expansion timescales (point type) and initial neutron excesses (color).
%The represented timescales are: $\tau_\mbox{decay}=0.1$~s (crosses), 0.2~s (filled circles), and
%0.6~s (triangles). The initial neutron excesses are $\eta=0.227\,10^{-3}$ (red),
%$\eta=0.909\,10^{-3}$ (blue) $\eta=6.818\,10^{-3}$ (black).}
%\label{fig13}
%\end{figure}

\subsubsection{Understanding the link between neutron excess and the final ratio of manganese to
chromium}\label{understanding1}

It is instructive to try to understand the origin of the tight dependence of the Mn/Cr ratio in
the X-ray epoch on the initial neutron excess. For this purpose, it is necessary to tell the
story of the synthesis of their parent species, \element[][55]{Co} and \element[][52]{Fe} (see
Table~\ref{tab1}). 

\begin{figure*}[tb]
\centering
   \includegraphics[width=9 cm]{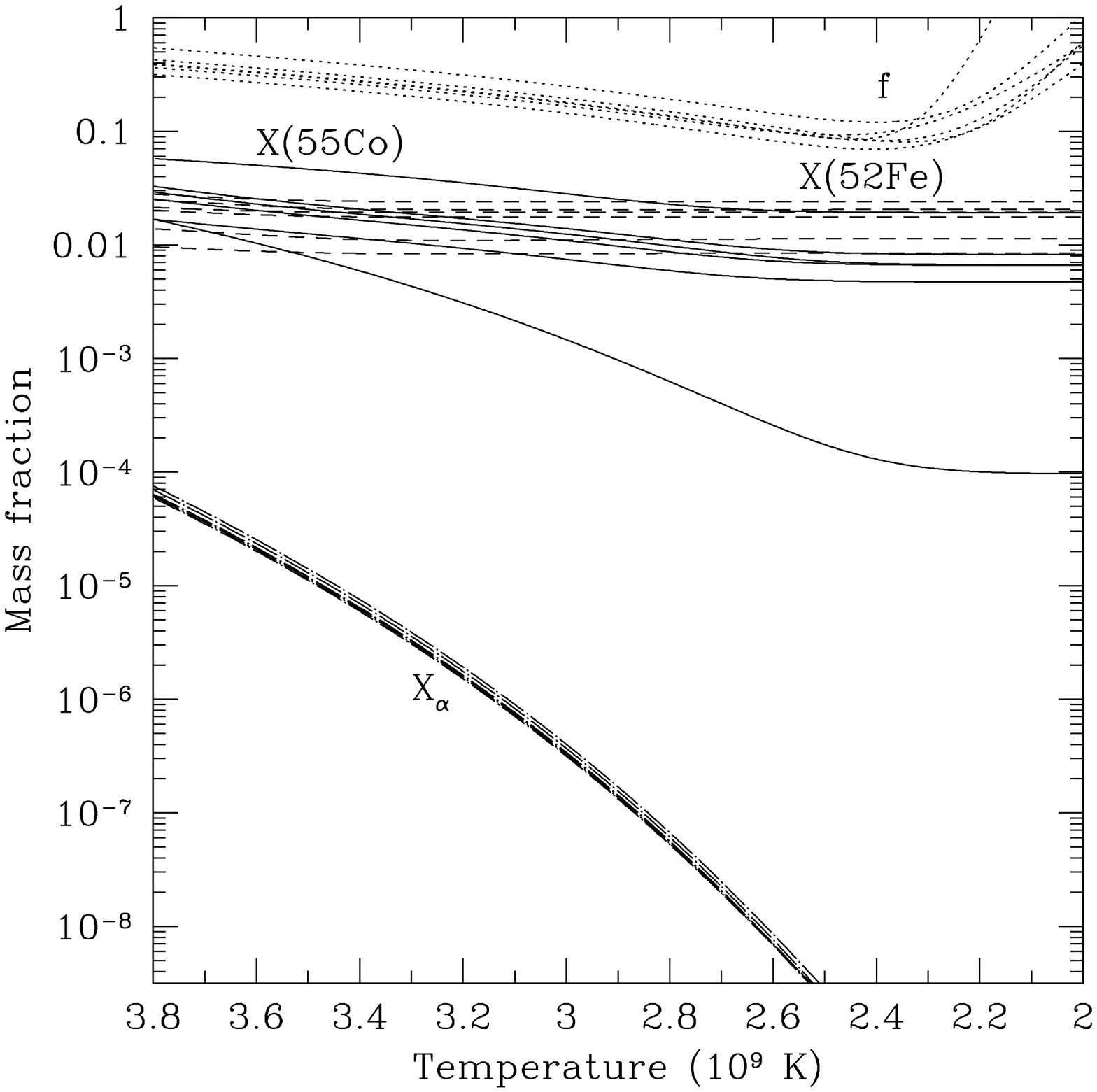}  \includegraphics[width=9 cm]{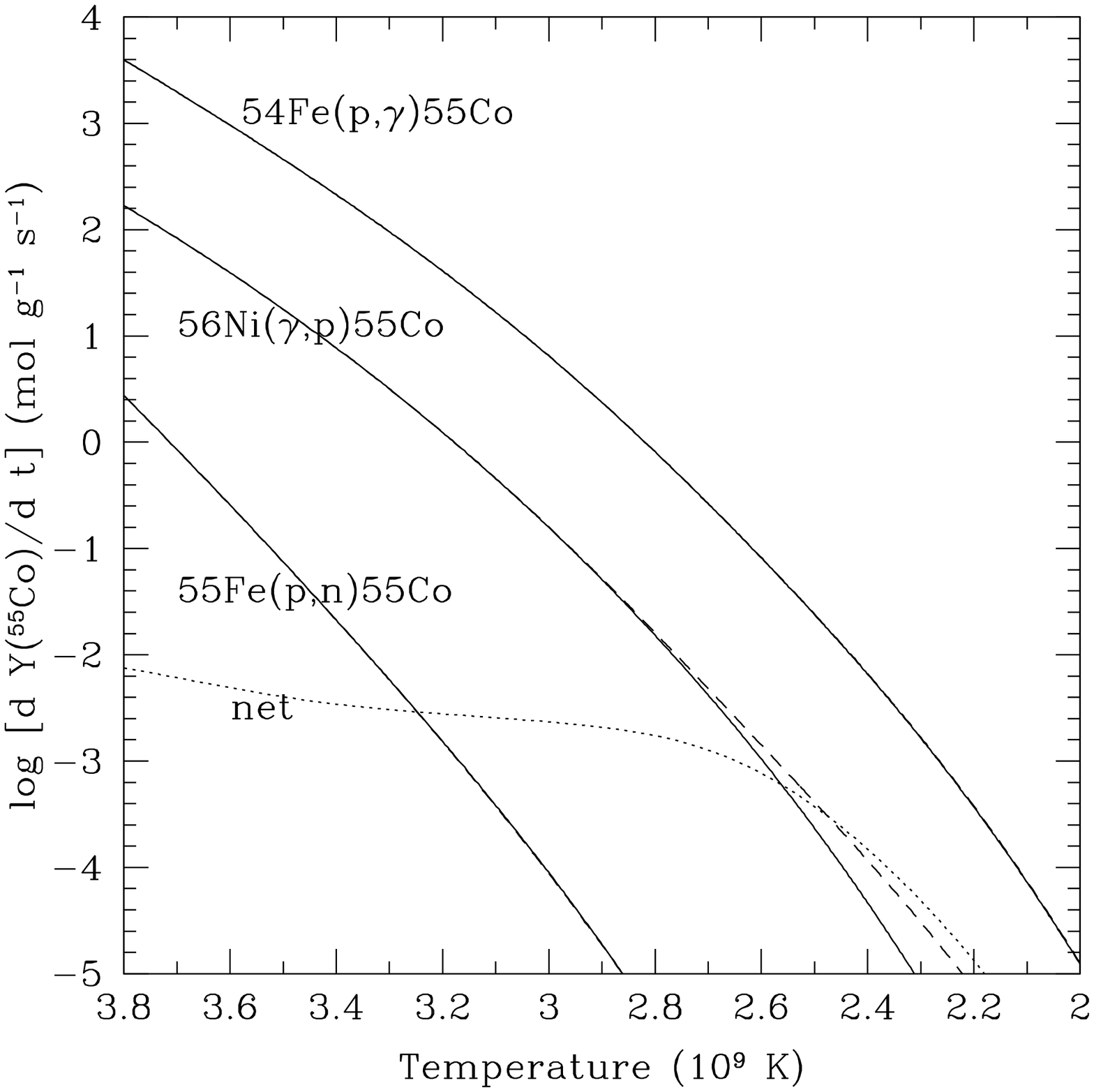}
\caption{Nuclear evolution as a function of temperature during expansion and cooling.
{\bf Left:} Evolution of the mass fractions of \element[][55]{Co} (solid lines), \element[][52]{Fe}
(dashed lines),
and $\alpha$ (dot-dashed lines), for six different sets of parameters, as given in
Table~\ref{tab2}. The dotted lines show the quantity $f$, defined in Eq.~\ref{eqdeff}.
{\bf Right:} Evolution of the molar fluxes of the main reactions contributing
to the abundance of \element[][55]{Co} for the same set of parameters as in Fig.~\ref{fig2}.
Reactions building \element[][55]{Co} are drawn as solid lines, reactions destroying it are drawn
as dashed
lines, and the absolute value of the net rate of $\mbox{d} Y(\element[][55]{Co})/\mbox{d} t$ is
drawn as a dotted line. Upper curves are for the reactions
$\element[][54]{Fe}+\mbox{p}\leftrightarrows\element[][55]{Co}+\gamma$, in which both line types
are
nearly indistinguishable. The middle curves are for the reactions
$\element[][56]{Ni}+\gamma\leftrightarrows\element[][55]{Co}+\mbox{p}$. Finally, the bottom curves
are for the reactions $\element[][55]{Fe}+\mbox{p}\leftrightarrows\element[][55]{Co}+\mbox{n}$.
}
\label{fig10}
\end{figure*}

At the peak temperatures studied here, both \element[][55]{Co} and \element[][52]{Fe} attain
abundances in statistical equilibrium
with the iron QSE-group. As temperature drops, both nuclei evolve to maintain this
equilibrium, which is broken when $T<3\times10^9$~K \citep{hix99}. The left hand panel of
Fig.~\ref{fig10} shows the evolution of the abundances of \element[][55]{Co} and \element[][52]{Fe}
during expansion
and cooling for six different sets of parameters, given in Table~\ref{tab2}. It can be seen that
the abundance of \element[][55]{Co} levels off at $T\simeq2.5\times10^9$~K, whereas that of
\element[][52]{Fe}
stabilizes even earlier, at $T\simeq3.5\times10^9$~K. The highest differences between the curves
belonging to \element[][55]{Co} are found for the extremes metallicities tested: the
bottom solid
curve, $X(\element[][22]{Ne})=2.5\times10^{-4}$, and the top solid curve, 
$X(\element[][22]{Ne})=0.075$. The abundance of \element[][55]{Co} remains in equilibrium with
theirs neighbors in the iron QSE-group until $T\simeq2\times10^9$~K, as can be seen in the right
hand panel of Fig.~\ref{fig10}. In this figure, the molar fluxes of the main
reactions linking \element[][55]{Co} with other members of the iron QSE-group are plotted,
where the most
important are $\element[][54]{Fe}+\mbox{p}\leftrightarrows\element[][55]{Co}+\gamma$ and
$\element[][56]{Ni}+\gamma\leftrightarrows\element[][55]{Co}+\mbox{p}$. The flux rate of the proton
capture on \element[][55]{Co} begins to depart from the rate of the $(\gamma,\mbox{p})$ reaction on
\element[][56]{Ni} just after $T\simeq2.8\times10^9$~K. However, the net rate of destruction of 
\element[][55]{Co} is always at least an order of magnitude lower than the
rate of each of the reactions
$\element[][54]{Fe}+\mbox{p}\leftrightarrows\element[][55]{Co}+\gamma$, thus ensuring that
equilibrium is maintained in the temperature range shown in Fig.~\ref{fig10}. Assuming that the
abundances within the iron QSE-group remain
in equilibrium allows us to formulate a
simple model for the ratio of the final (before radioactive disintegrations)
mass fractions of \element[][55]{Co} to \element[][52]{Fe},
$X(\element[][55]{Co})/X(\element[][52]{Fe})$. 

\begin{table}[tb]
\caption{Sets of parameters of Si-b represented in the left panel of Fig.~\ref{fig10}.}\label{tab2}
\centering
\begin{tabular}{lllll} 
\hline\hline             
$T_{\mbox{9,peak}}$ & $\tau$ & $X(\element[][22]{Ne})$ & $\tau_{\mbox{rise}}$ &
$X(\element[][12]{C})$
\\
& (s) & & (s) & \\
\hline
4.8 & 0.29 & 0.010 & $10^{-9}$ & 0.5 \\
4.3 & 0.29 & 0.010 & $10^{-9}$ & 0.5 \\
5.2 & 0.29 & 0.010 & $10^{-9}$ & 0.5 \\
4.8 & 0.72 & 0.010 & $10^{-9}$ & 0.5 \\
4.8 & 0.29 & 0.075 & $10^{-9}$ & 0.5 \\
4.8 & 0.29 & $2.5\times10^{-4}$ & $10^{-9}$ & 0.5 \\
\hline
\end{tabular}
\end{table}

We begin by assuming that the initial neutron excess in the matter subject to incomplete Si-b
is stored in the species \element[][54]{Fe} and \element[][55]{Co} at the end of the
nucleosynthetic epoch, which can be justified in light of
Figs.~\ref{fig2} and Figs.~\ref{fig6} - \ref{fig8}. Then, using Eq.~\ref{eqeta}, one
can write
\begin{equation}\label{eq5455}
 2Y_{22,0} = 2Y_{54}+Y_{55}\,,
\end{equation}
\noindent where $Y_{22,0}=X(\element[][22]{Ne})/22$ is the initial molar fraction of
\element[][22]{Ne} and
for
simplicity, we write the molar fraction of species $^{A}Z$ as $Y_{A}$, in mol/g.
%(at this point, it should be clear to which elements belong each barion number $A$). 
Another relationship between the
abundances of \element[][54]{Fe} and \element[][55]{Co} can be derived from the assumption of
equilibrium of the
forward and backward $(\mbox{p},\gamma)$ reactions linking both nuclei, and the ones linking
\element[][55]{Co} with \element[][56]{Ni},
\begin{equation}\label{eqequil1}
 Y_{55} = \left( Y_{54} Y_{56} \frac{r_{54\mbox{p}\gamma}\lambda_{56\gamma\mbox{p}}}
{r_{55\mbox{p}\gamma}\lambda_{55\gamma\mbox{p}}} \right)^{1/2} 
 = \left[ Y_{54} Y_{56} \frac{C(\element[][55]{Co})^2}{C(\element[][54]{Fe}) C(\element[][56]{Ni})}
\right]^{1/2}\,,
\end{equation}
\noindent where $r_{54\mbox{p}\gamma}$ is the rate of radiative proton capture on
\element[][54]{Fe},
$\lambda_{56\gamma\mbox{p}}$ is the rate of photodisintegration of \element[][56]{Ni} through the
proton
channel, and so on. The last equality is a result of the assumption of global equilibrium
within the iron QSE-group, where the quantity $C(^AZ)$ is defined by \citep[see, e.g.][]{hix96},
\begin{equation}
 C(^AZ) = \frac{G(^AZ)}{2^A} \left(\frac{\rho N_{\mbox{A}}}{\theta}\right)^{A-1} A^{3/2} \exp
\left[\frac{B(^AZ)}{kT}\right]\,,
\end{equation}
\noindent where $G(^AZ)$ is the partition function, $B(^ZA)$ the nuclear binding energy,
$N_{\mbox{A}}$ Avogadro's number, $k$ Boltzmann's constant, and
$\theta=5.9417\times10^{33}T_9^{3/2}$. 

Solving Eq.~\ref{eqequil1} for $Y_{54}$ and substituting in Eq.~\ref{eq5455} gives the following
relationship between the final abundance of \element[][55]{Co} and the initial abundance of
\element[][22]{Ne},
\begin{equation}\label{eq55}
 Y_{55} = \frac{\mathfrak{A}^2}{4}\left(\sqrt{1+\frac{16Y_{22,0}}{\mathfrak{A}^2}}-1\right)\,,
\end{equation}
\noindent where $\mathfrak{A}$ has been defined as
\begin{equation}
 \mathfrak{A} \equiv 8Y_{56}^{1/2} \exp\left(-\frac{12.19}{T_9}\right)\,.
\end{equation}
\noindent Equation~\ref{eq55} sets the functional dependence of $Y_{55}$ on
$X(\element[][22]{Ne})$ (or $\eta$): for $X(\element[][22]{Ne})\ll\mathfrak{A}^2$ it leads to $Y_
{55}\propto X(\element[][22]{Ne})$, whereas for $X(\element[][22]{Ne})\gg\mathfrak{A}^2$ the result
is $Y_
{55}\propto X(\element[][22]{Ne})^{1/2}$. Since $\mathfrak{A}\sim 0.011$ when global QSE breaks
down
($T_9\simeq 2.9$), it turns out that we are in an intermediate regime, in agreement with the
power-law exponent in Eq.~\ref{eqfitmncr}.

In QSE, the molar abundance of \element[][52]{Fe} can be written as
\begin{equation}
 Y_{52}= \frac{Y_{56}}{Y_{\alpha}} \frac{C(\element[][52]{Fe}) C(\alpha)}{C(\element[][56]{Ni})}\,,
\end{equation}
\noindent From this equation and Eq.~\ref{eq55} one obtains
\begin{equation}
 \frac{X(\element[][55]{Co})}{X(\element[][52]{Fe})} = f
\left(\sqrt{1+\frac{16Y_{22,0}}{\mathfrak{A}^2}}-1\right)\,,
\end{equation}
\noindent where
\begin{equation}\label{eqdeff}
 f\equiv 2.4\times10^{-10} \frac{\rho Y_{\alpha}}{T_9^{3/2}} \exp\left(\frac{68.4}{T_9}\right)\,
\end{equation}
\noindent is shown in Fig.~\ref{fig10} to only vary slightly for $3.8 > T_9 \ga 2.2$ and to be
nearly independent of the parameters of the Si-b nucleosynthetic calculation. Substituting
approximate values for $f\simeq 0.07$ and $\mathfrak{A}\simeq 0.011$, one finally obtains the
desired relationship between the abundances of \element[][55]{Co} and \element[][52]{Fe}, and the
neutron excess, 
\begin{equation}\label{eqanamncr}
 \frac{X(\element[][55]{Co})}{X(\element[][52]{Fe})} \simeq 0.07
\left(\sqrt{1+6.61\times10^4\eta}-1\right)\,.
\end{equation}
\noindent This equation nicely fits the results of the nucleosynthesis calculations, as can be
seen in Fig.~\ref{fig9}. We stress that the approximations needed to derive Eq.~\ref{eqanamncr}
imply that it is not necessarily better than the simpler power-law fit given by
Eq.~\ref{eqfitmncr}, which is shown in the same figure. However, its derivation allows us to
discern where
the dependence of $M(\mbox{Mn})/M(\mbox{Cr})$ on the progenitor metallicity comes
from, and what its conditions of validity are. It also shows that any effect of the rest of
parameters is of
second order in comparison with that related to a variation in $Z$. Finally, it also allows us to
understand that the
power-law index of 0.676 is the result of the comparable magnitude of $X(\element[][22]{Ne})$ and
$\mathfrak{A}$.

\subsection{Ratio of vanadium to titanium}                       

\begin{figure*}[tb]
\centering
   \includegraphics[width=9 cm]{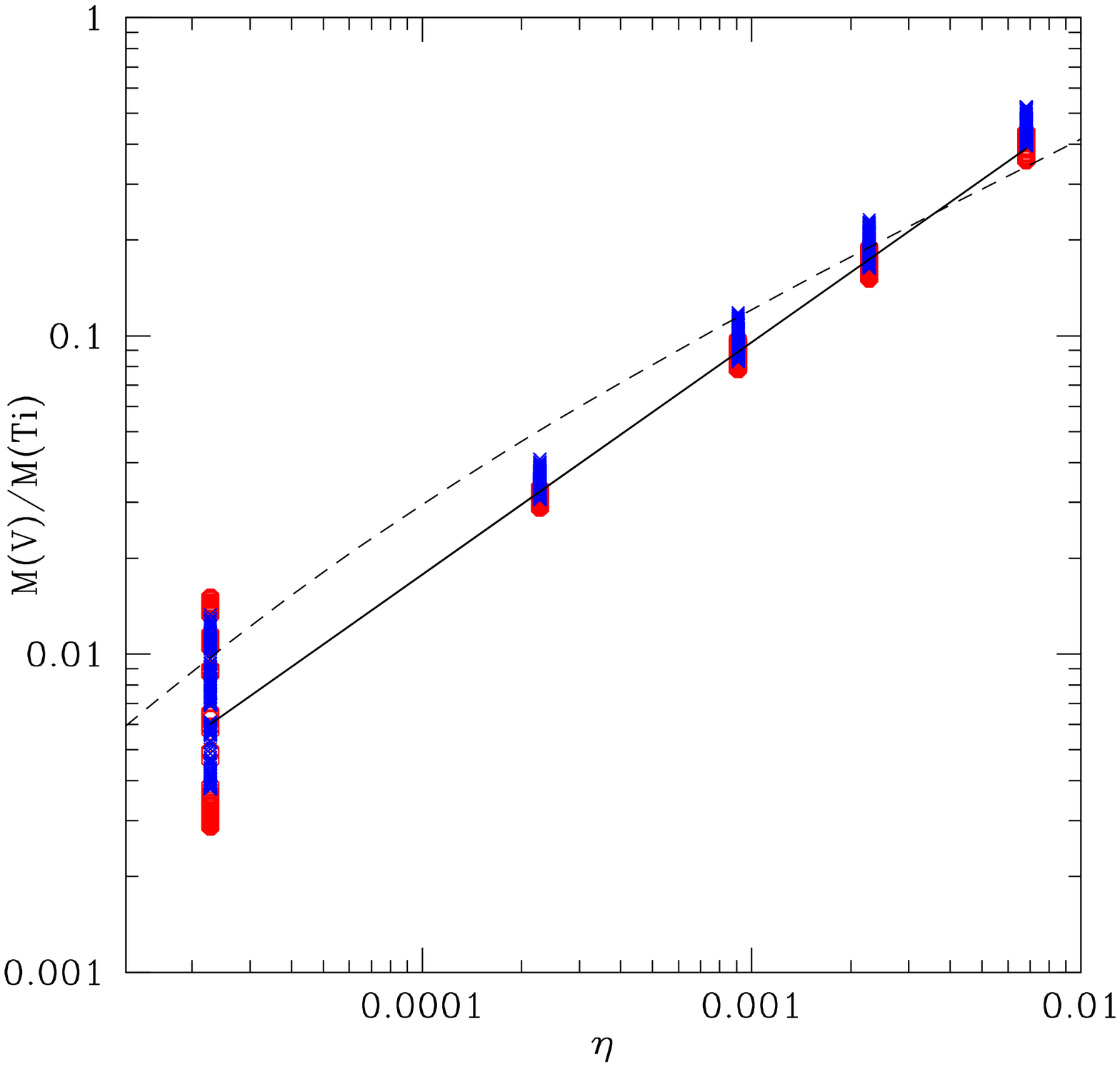}  \includegraphics[width=9 cm]{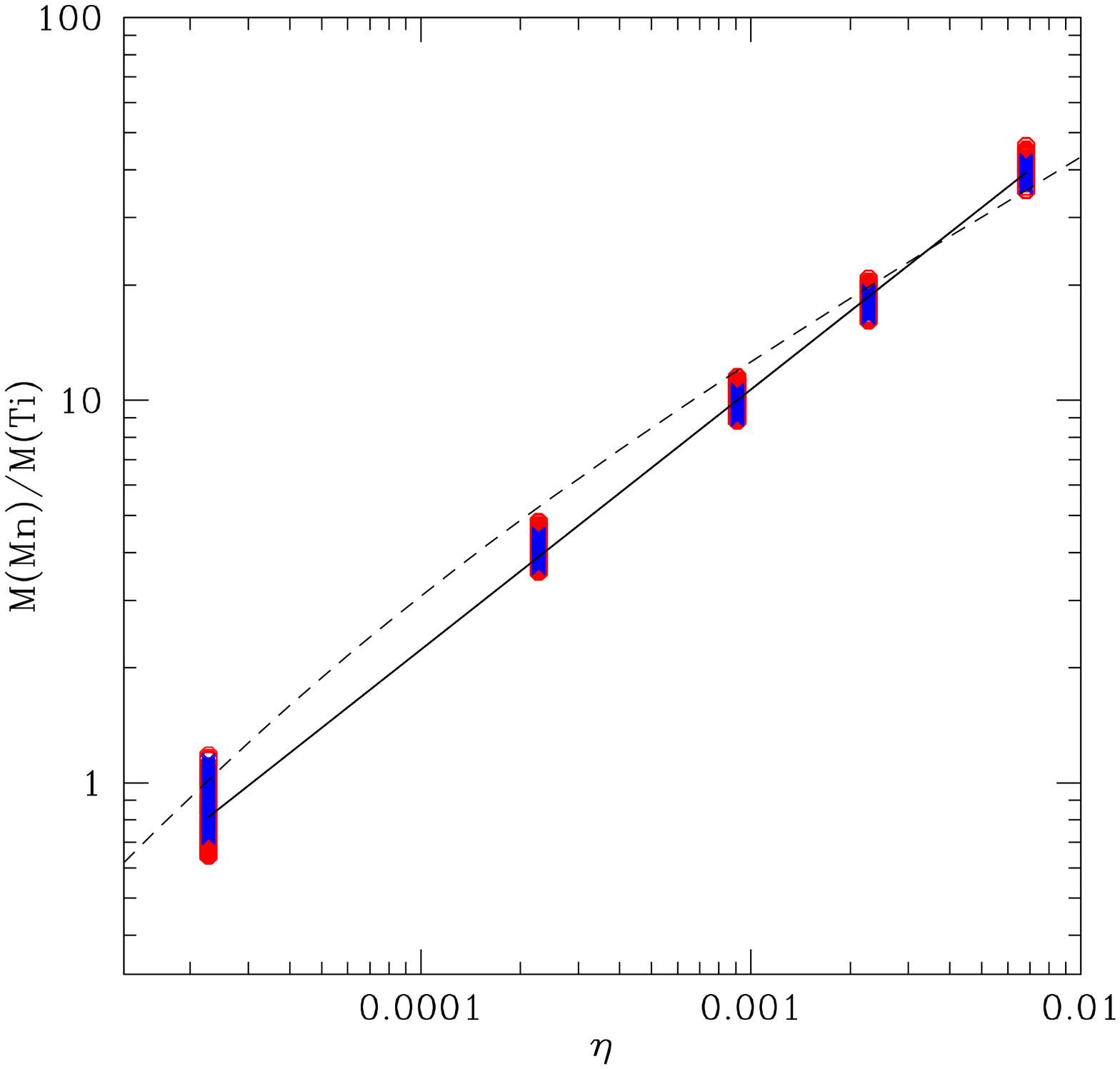}
\caption{(Color online)
Averaged final mass fraction ratios of vanadium to titanium and of manganese to titanium as tracers
of the initial neutron excess. The meaning of the point types and colors, as well as the ranges of 
parameters are the same as in Fig.~\ref{fig9}.
{\bf Left:} Mass ratio  V/Ti. The solid line is a fit given by,
$M(\mbox{V})/M(\mbox{Ti})=14.8\eta^{0.73}$.
The dashed line shows the relationship expected from an analytic model, Eq.~\ref{eqanavti}.
{\bf Right:} Mass ratio  Mn/Ti. The solid line is a fit given by,
$M(\mbox{Mn})/M(\mbox{Ti})=1170\eta^{0.68}$.
The dashed line shows the relationship expected from an analytic model, Eq.~\ref{eqanamnti}.
}
\label{fig11}
\end{figure*}

The ratio of abundances of vanadium to titanium, V/Ti, provides an alternative way of measuring the
initial neutron excess of the progenitor of a SNIa in
the X-ray phase. The left hand panel of Fig.~\ref{fig11} shows that 
V/Ti traces nicely the initial neutron excess apart from $\eta\la 10^{-4}$, for which there is a
substantial 
dispersion of this ratio. The best power-law fit is
\begin{equation}\label{eqfitvti}
   M(\mbox{V})/M(\mbox{Ti})=14.8\eta^{0.726}\,.
\end{equation}
\noindent Notice the similarity between the exponent of the power laws in Eqs.~\ref{eqfitmncr} and
\ref{eqfitvti}.
According to Fig.~\ref{fig1}, the ratio V/Ti obtained with the analytic model of Si-b given by
Eqs.~\ref{eqT} and 
\ref{eqro} matches the ratio  in hydrodynamical simulations of SNIa very well.

In the X-ray epoch, vanadium and titanium come from \element[][51]{Mn} and \element[][48]{Cr}, 
respectively (see Table~\ref{tab1}), whose abundances in the nucleosynthesis epoch are linked by
$\alpha$ 
captures to those of \element[][55]{Co} and \element[][52]{Fe}. Using the equilibrium
relationships, as in 
section~\ref{understanding1}, the abundance ratio of \element[][51]{Mn} to \element[][48]{Cr} can
be shown to be proportional to the ratio of \element[][55]{Co} to \element[][52]{Fe}, with
%\begin{equation}\label{eqanavti}
% \frac{X(^{51}\mbox{Mn})}{X(^{48}\mbox{Cr})} \simeq 
%0.76 \exp\left(-\frac{3.146}{T_9}\right) \frac{X(^{55}\mbox{Co})}{X(^{52}\mbox{Fe})} \simeq
%\left(0.15 - 0.30\right) \frac{X(^{55}\mbox{Co})}{X(^{52}\mbox{Fe})}\,,
%end{equation}
\begin{eqnarray}\label{eqanavti}
  \frac{X(\element[][51]{Mn})}{X(\element[][48]{Cr})} & \simeq 
0.76 \exp\left(-\frac{\displaystyle 3.146}{\displaystyle T_9}\right)
\frac{\displaystyle X(\element[][55]{Co})}{\displaystyle X(\element[][52]{Fe})} \nonumber\\
 & \simeq \left(0.15 - 0.30\right) \frac{\displaystyle
X(\element[][55]{Co})}{\displaystyle X(\element[][52]{Fe})} \,;
\end{eqnarray}
\noindent i.e., the abundance ratio  V/Ti is expected to range about 15-30\% of the ratio  Mn/Cr.
Comparison 
of the left hand panel of Figs.~\ref{fig11} to \ref{fig9} confirms this approximate
relationship. 

\subsection{Ratio of manganese to titanium}

The ratio of abundances of manganese to titanium, Mn/Ti, provides an alternative way of
measuring the initial neutron excess in the X-ray phase. The right hand panel of Fig.~\ref{fig11}
shows that Mn/Ti nicely traces the initial neutron excess in the whole $\eta$ range explored here.
The best power-law fit is
\begin{equation}\label{eqfitmnti}
   M(\mbox{Mn})/M(\mbox{Ti})=1170\eta^{0.677}\,,
\end{equation}
\noindent which is again similar to the exponents of the power laws in Eqs.~\ref{eqfitmncr} 
and \ref{eqfitvti}.
According to Fig.~\ref{fig1}, the ratio Mn/Ti obtained with the analytic model of Si-b given by
Eqs.~\ref{eqT}
and \ref{eqro} shows a relative error of approximately 20\% relative to the ratio  in
hydrodynamical 
simulations of SNIa. However, this error is much lower than the range of variation in 
$M(\mbox{Mn})/M(\mbox{Ti})$ as a function of $\eta$.

In the X-ray epoch, manganese and titanium come from \element[][55]{Co} and \element[][48]{Cr}, 
respectively (see Table~\ref{tab1}). Using the equilibrium relationships,
the ratio of \element[][55]{Co} to \element[][48]{Cr} can be shown, as in the previous
section, to be proportional to the ratio
of \element[][55]{Co} to \element[][52]{Fe}, with
\begin{eqnarray}\label{eqanamnti}
 \frac{X(\element[][55]{Co})}{X(\element[][48]{Cr})} & \simeq 
1.43\times10^{-11} \frac{\displaystyle \rho}{\displaystyle T_9^{3/2}}
\exp\left(\frac{\displaystyle 92.15}{\displaystyle T_9}\right) Y_{\alpha} 
\frac{\displaystyle X(\element[][55]{Co})}{\displaystyle X(\element[][52]{Fe})} \nonumber \\
 & \simeq 25 \frac{\displaystyle X(\element[][55]{Co})}{\displaystyle X(\element[][52]{Fe})} \,. 
\end{eqnarray}
\noindent Comparison of the right hand panel of Figs.~\ref{fig11} to \ref{fig9} confirms this
approximate relationship. 

\section{Tracers of the progenitor metallicity: Optical epoch}

As seen in Table~\ref{tab1}, there are several important contributors to the abundances of the
elements from scandium to iron in the optical phase, most of them unstable, which implies that
their abundances vary in time. The variation in time of their ratios makes it more
difficult to define in this phase a tracer of the explosion properties. However, there are
interesting correlations between the progenitor metallicity and several abundance ratios in
the optical phase, namely manganese to chromium, vanadium to manganese, and titanium to manganese.
All
these ratios share the need to measure the abundance of manganese. We have not been
able to found any tracer of the progenitor metallicity in this phase that does not involve
manganese.

\begin{figure}[tb]
\centering
   \includegraphics[width=9 cm]{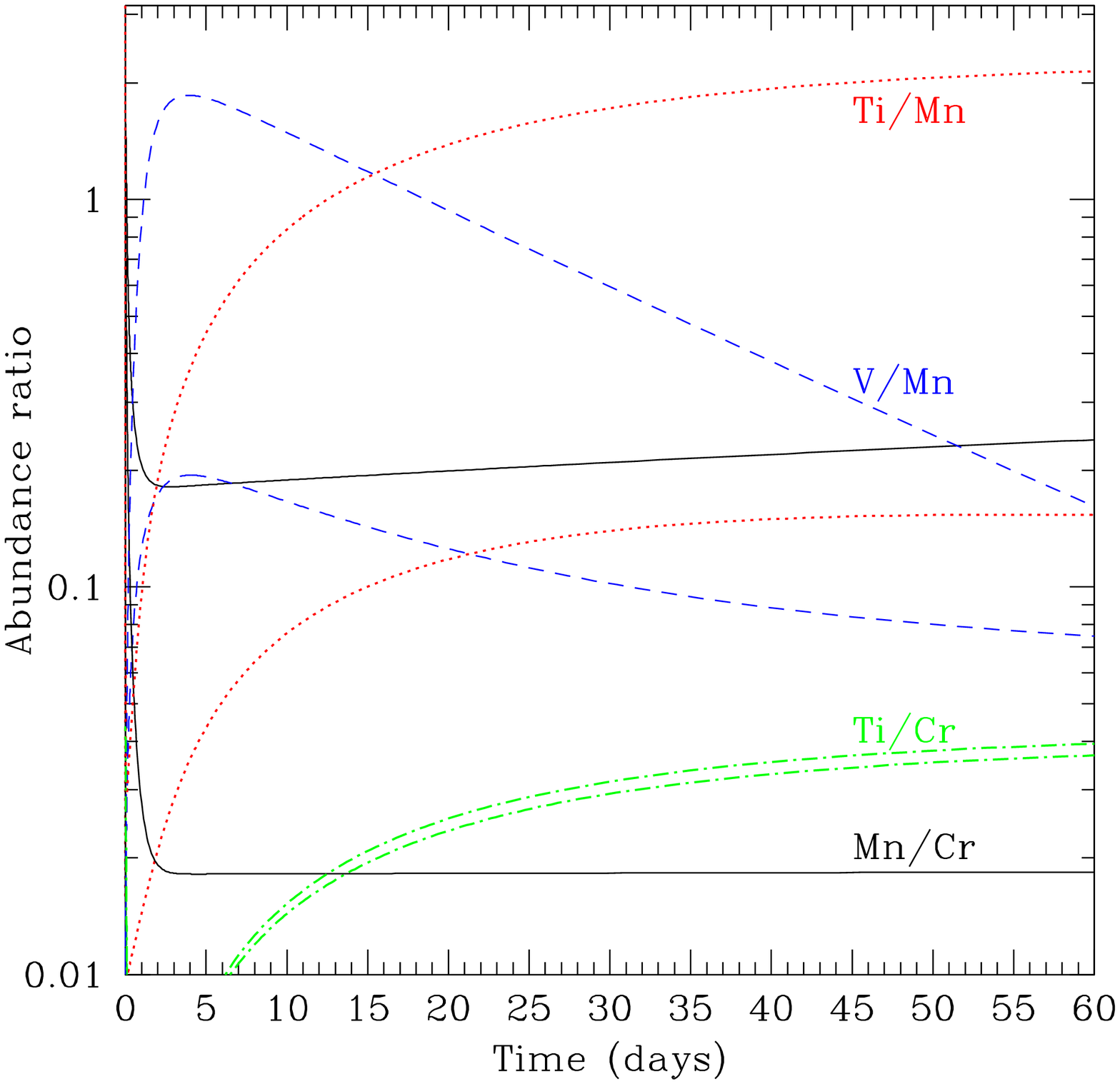}
\caption{(Color online)
Evolution with time of the abundance ratios of potential tracers of the progenitor metallicity
in the optical epoch: solid black lines belong to Mn/Cr, dashed blue lines to V/Mn, and dotted
red lines to Ti/Mn. For each couple of elements, the labeled curve is the one belonging to
$X(\element[][22]{Ne})=2.5\times10^{-4}$, whereas the other curve with the same type and color
belongs
to $X(\element[][22]{Ne})=0.075$. The dot-dashed green line shows the evolution of Ti/Cr ratio,
which could be used as a control variable, see section~\ref{control}.
}
\label{fig12}
\end{figure}

Figure~\ref{fig12} shows the evolution of several abundance ratios for the extreme values of the
initial neutron excess explored in the present
work. The separation between the curves belonging to the different $\eta$ makes these abundance
ratios promising tracers of the initial neutron excess.

\subsection{Ratio of manganese to chromium}

Among the element couples shown in Fig.~\ref{fig12}, the ratio
Mn/Cr exhibits the best behavior, because it remains nearly constant after day three.
At these times, manganese is composed of \element[][53]{Mn}, made as \element[][53]{Fe} in the
nucleosynthetic epoch, in a proportion higher than 80\%, and the dominant isotope of chromium is
\element[][52]{Cr}, 
made as \element[][52]{Fe}, which represents more than 90\% of its abundance. 

\begin{figure}[tb]
\centering
   \includegraphics[width=9 cm]{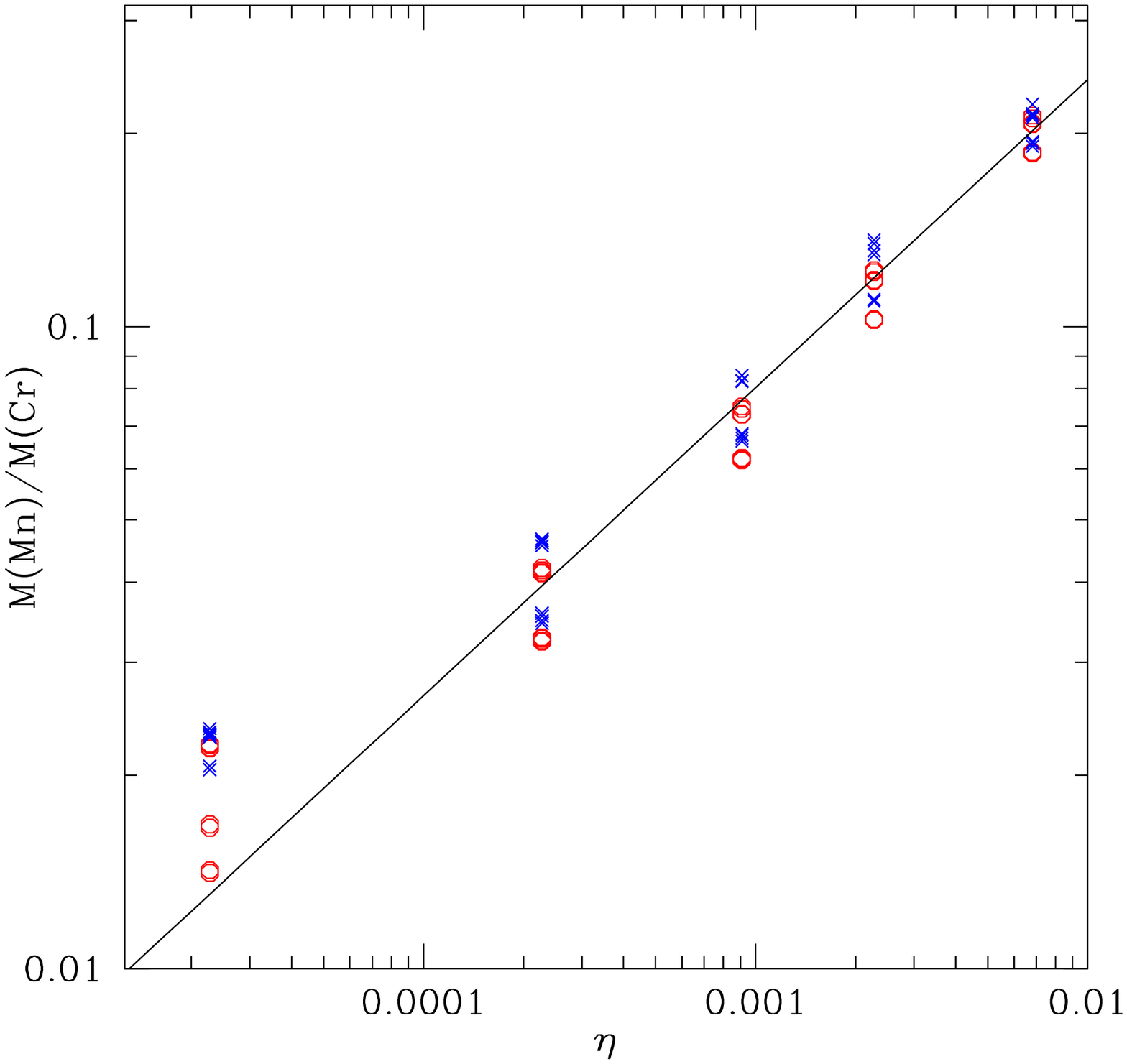}
\caption{(Color online)
Averaged final mass fraction ratio  Mn/Cr in the optical phase, as a function of the initial
neutron excess.
The meaning of the point types and colors is the same as in Fig.~\ref{fig9}. The ranges of the
parameters $\tau_{\mbox{rise}}$, $\tau$, and $X(\element[][12]{C})$ are  as in
Fig.~\ref{fig9}. The solid line is a fit given by,
$M(\mbox{Mn})/M(\mbox{Cr})=2.2\eta^{0.48}$.
}
\label{fig13}
\end{figure}

As can be seen in Fig.~\ref{fig13}, the ratio Mn/Cr in the optical phase is a good tracer of
the initial neutron excess. The dispersion related to the different parameters used in the present
calculations of Si-b is small, although it is larger for the lowest initial neutron excess. The
best power-law fit is
\begin{equation}\label{eq22}
 M(\mbox{Mn})/M(\mbox{Cr})=2.2\eta^{0.48}\,.
\end{equation}

\subsection{Ratio of vanadium to manganese}

At the optical epoch, there are various vanadium isotopes contributing significantly to the
element abundance, namely \element[][48]{V}, \element[][49]{V},
and \element[][51]{V}. Isotope \element[][48]{V}, whose parent decays in less than one day,
disintegrates itself
with a lifetime slightly less than the time it takes for a typical SNIa to reach maximal
brightness. Isotope \element[][49]{V}, formed from \element[][49]{Cr} practically instantaneously
after the
explosion, has a lifetime close to one year, which makes its contribution close to constant
in the whole optical epoch. Isotope \element[][51]{V} also forms practically instantaneously after
the
explosion from \element[][51]{Mn}, and decays with a lifetime of nearly one month. At the
lowest
neutron excess we considered, only \element[][48]{V} contributes to the vanadium abundance. 

\begin{figure*}[tb]
\centering
   \includegraphics[width=9 cm]{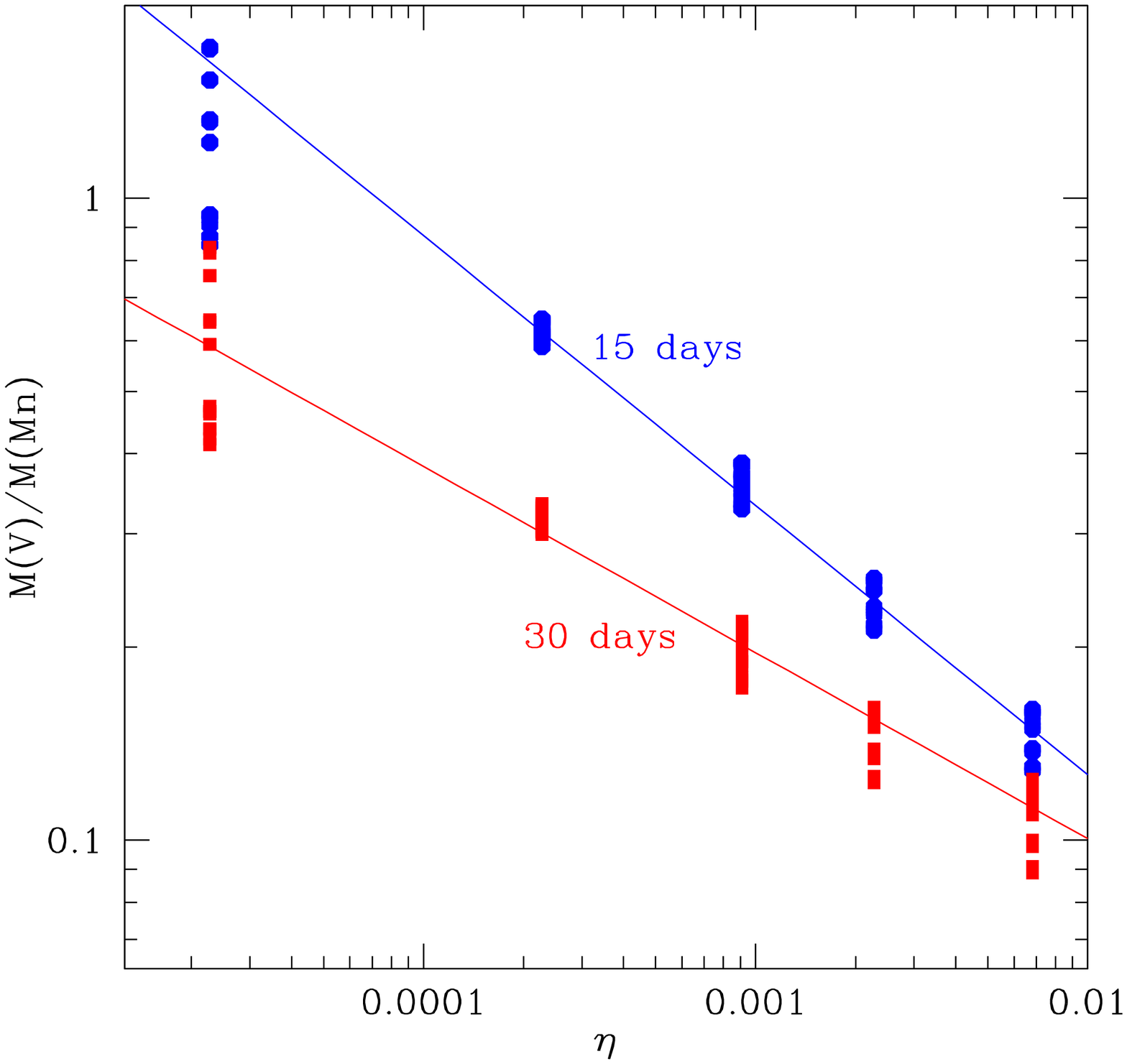} \includegraphics[width=9 cm]{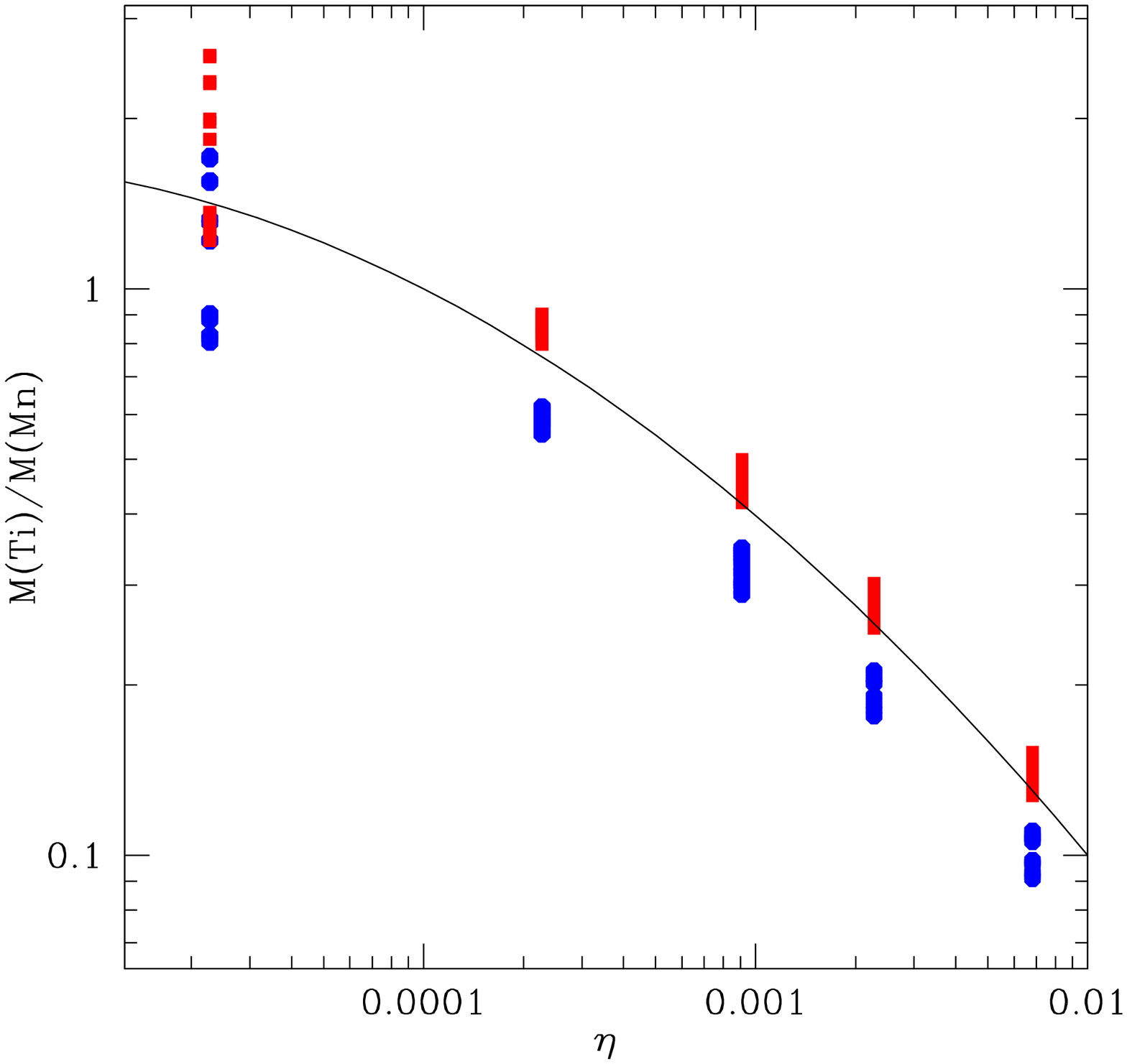}
\caption{(Color online)
Averaged final mass fraction ratios V/Mn and Ti/Mn in the optical phase, as a function of the
initial neutron excess.
The point types and colors in these plots refer to the time at which the ratios are
drawn, blue solid circles are calculated at day 15, whereas red solid squares belong to day 30. The
ranges of the parameters $\tau_{\mbox{rise}}$, $\tau$, and $X(\element[][12]{C})$ are as in
Fig.~\ref{fig9}. 
{\bf Left:} Abundance ratio of vanadium to manganese. The ratios are fit by power-law
functions, namely $M(\mbox{V})/M(\mbox{Mn})=0.018/\eta^{0.42}$ for the points at day 15, and
$M(\mbox{V})/M(\mbox{Mn})=0.026/\eta^{0.29}$ for the points at day 30.
{\bf Right:} Abundance ratio of titanium to manganese. In this case, the points belonging to days
15 and 30 can be reasonably fit by a single function, although it cannot be a simple power law.
The fit function shown is, $M(\mbox{Ti})/M(\mbox{Mn})=\exp\left(-6.45 - 1.1\ln\eta -
0.043\ln^2\eta\right)$.
}
\label{fig14}
\end{figure*}

The abundance ratio of vanadium to manganese is another good tracer of the initial neutron excess,
although it presents the difficulty that the ratio changes with time. The left hand panel of
Fig.~\ref{fig14} shows this ratio and a power-law fit for two times. At 15 days, close to maximal
brightness of a typical SNIa, it is given by
\begin{equation}\label{eq23}
 M(\mbox{V})/M(\mbox{Mn})=0.018/\eta^{0.42}\,,
\end{equation}
\noindent
whereas at 30 days, a time at which the photosphere is close to the center
of the ejecta, the fit is
\begin{equation}\label{eq24}
 M(\mbox{V})/M(\mbox{Mn})=0.026/\eta^{0.29}\,.
\end{equation}

\subsection{Ratio of titanium to manganese}

The dominant isotope of titanium after day 10 is
\element[][48]{Ti}, made as \element[][48]{Cr}, whose abundance amounts to more than 90\% of the
total mass of
titanium. Although the lifetime of the radioactive parent of \element[][48]{Ti}, i.e.
\element[][48]{V}, is 16 days,
the ratio Ti/Mn changes little after day $\sim15$ in comparison with its dependence on the initial
neutron excess. The right hand panel of Fig.~\ref{fig14} shows that Ti/Mn can be fit by the same
function of $\eta$ at 15 and 30 days,
\begin{equation}\label{eq25}
 M(\mbox{Ti})/M(\mbox{Mn})=\exp\left(-6.45 - 1.1\ln\eta - 0.043\ln^2\eta\right)\,.
\end{equation}

\section{Tracers of the explosion conditions}

As seen in the previous section, the ratios of abundances of most iron QSE-group elements are
especially sensitive to the neutron excess. Consequently, it is difficult to find tracers of the
rest of parameters, namely expansion timescale, temperature rise time, and initial carbon
abundance. We have
 only found one such possible tracer, V/Mn in the X-ray epoch.

\subsection{Expansion timescale: Ratio of vanadium to manganese at the X-ray epoch}

The most promising pairs of elements to look for a tracer of the expansion timescale, $\tau$, are
those that come from parents linked by an alpha capture in QSE, because this is the reaction type
that depends the less in the n/p ratio, hence in the neutron excess. 
The first possibility is the pair formed by titanium and chromium, at the X-ray epoch, whose
grandparents in the nucleosynthetic epoch are \element[][48]{Cr} and \element[][52]{Fe},
respectively. However, the
results of the present calculations show that the ratio of their abundances is insensitive to
$\tau$.

The second possibility is the pair formed by vanadium and manganese, also at the X-ray epoch,
whose grandparents are \element[][51]{Mn} and \element[][55]{Co}, respectively. Their abundance
ratio is only mildly
sensitive to $\tau$ and strongly dependent on $\eta$. An approximate measure of the
expansion timescale would only be possible once the neutron excess is known, and only for
low-metallicity progenitors, $Z\la 0.1Z_{\sun}$. At higher values of the initial neutron excess,
the ratio of vanadium to manganese abundances, V/Mn, is insensitive to $\tau$. 

\begin{figure}[tb]
\centering
   \includegraphics[width=9 cm]{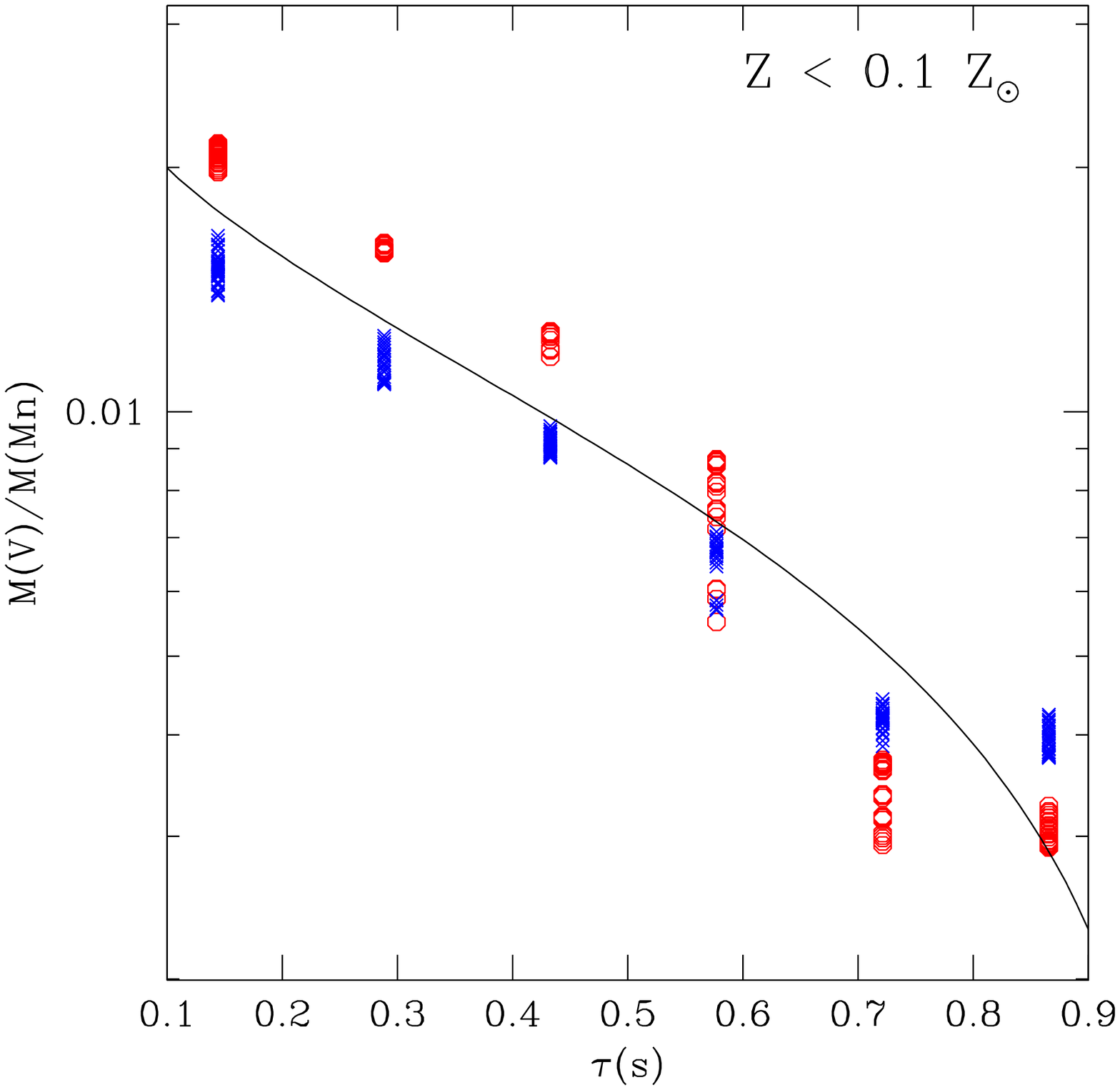}
\caption{(Color online)
Averaged final mass fraction ratio of vanadium to manganese as tracer
of the expansion timescale, at fixed neutron excess, $\eta=2.27\times10^{-5}$. The meaning of the
point types and colors is the same as in Fig.~\ref{fig9}. The results are
shown for different $\tau_{\mbox{rise}}$ in the range $10^{-9}$-0.2~s, and
$X(\element[][12]{C})$ in the range 0.3-0.7. The
solid line is a fit given by 
$M(\mbox{V})/M(\mbox{Mn})=\log\left[\log\left(1/\tau\right)^{0.7}\right]-1.7$. 
}
\label{fig15}
\end{figure}

Figure~\ref{fig15} shows the ratio V/Mn for a low-metallicity progenitor as a function of the
expansion timescale. The calculated ratio depends on $\tau$ only for expansion timescales less
than $\sim0.8$~s. The dispersion introduced by the different peak-temperature distribution
functions used is comparable to the difference in V/Mn for close $\tau$, so this tracer would
not even allow determining the expansion timescale with high precision. The function,
\begin{equation}\label{eq26}
 M(\mbox{V})/M(\mbox{Mn})=\log\left[\log\left(1/\tau\right)^{0.7}\right]-1.7\,,
\end{equation}
\noindent provides a reasonable fit to the present results.

\subsection{Deflagration or detonation?}

Knowing the temperature rise timescale would allow discerning whether burning in the incomplete
Si-b regime in SNIa proceeds through a deflagration or a detonation wave. Unfortunately, as
can be expected from the discussion in section~\ref{secdefdet}, we did not found any pair of
iron QSE-group elements whose mass ratio is sensitive to the rise time. The same applies to the
determination of the initial carbon mass fraction.

\section{The ratio of titanium to chromium as a control variable}\label{control}

We have already explained that the ratio of the abundance of titanium to that of chromium, Ti/Cr,
is insensitive to the parameters of the incomplete Si-b calculation, in the limits explored in
the present work; i.e., it cannot be used as a tracer of any one of these parameters, including
the neutron excess. However, this does not mean that measuring Ti/Cr is useless. On the contrary,
it can be utilized as a control variable to check that the SNIa properties match the predictions
of the present Si-b calculations. 

Figure~\ref{fig12} shows the evolution of Ti/Cr in time until two months after a SNIa explosion.
Titanium abundance grows steadily as a result of the decay of \element[][48]{V}, whereas the
abundance of chromium remains stable after a few hours from the thermonuclear runaway. However,
the ratio Ti/Cr does not change significantly after day 30, and what is most important, it is
practically the same irrespective of the metallicity of the progenitor WD. In the X-ray epoch,
the full set of parameters used in the Si-b calculations leads to values of Ti/Cr in the range
$0.035 - 0.06$. 

The profile of Ti/Cr in the ejecta of the reference SNIa model of the present work, DDTc, is flat
in the region
undergoing incomplete Si-b ($4.3\le T_{\mbox{9,peak}}\le 5.2$) with a value in the range
$\mbox{Ti}/\mbox{Cr}\simeq0.025 - 0.05$, whereas it rises in the inner NSE region by nearly an
order of magnitude. As a result, measuring Ti/Cr might provide a way to evaluate whether the
observed signal from these elements comes from a region processed to incomplete Si-b, NSE, or a
mixture of both. 

\section{Discussion}

In this work, we investigate the properties of the nucleosynthesis in the incomplete Si-b regime, 
using a parameterized description of the initial chemical composition and the evolution of the
thermodynamic variables,
in the context of SNIa. We focus the study on potential nucleosynthetic tracers of the 
parameters of Si-b. We discarded tracers based on the measurement of the abundance of a single
element, 
because it would depend on the extent of combustion within the exploding WD, which is expected to
vary 
substantially from event to event. We therefore consider abundance ratios as the most promising
targets for 
the present study. We find that the ratios that involve elements belonging to the silicon
QSE-group do not
correlate well with the parameters of Si-b, so the selected ratios are those involving the
lightest 
elements of the iron QSE-group, i.e., titanium, vanadium, chromium, and manganese. We hereby
summarize the 
findings of this work and briefly discuss the plausibility of measuring the most
interesting 
abundance ratios either in SNIa remnants or in SNIa near maximum brightness, the so-called
photospheric epoch.

\subsection{Young supernova remnants}

A young supernova remnant is one in which the reverse shock, which is responsible for heating and
decelerating the 
ejecta, has not had time to reach the center of the ejecta. This time coincides more or less
with the moment when the shocked mass of the interstellar medium equals the supernova ejecta mass.
Since the X-ray emission of SNIa ejecta peaks at or near the 
reverse shock, it can reveal the chemical composition at different locations within the remnant.
Assuming 
chemical stratification of the ejecta, in the first thousand years or so after the explosion,
where the precise 
time depends on a number of environment and explosion variables, the emission comes from material
whose 
composition was set in the explosion by the incomplete Si-b process.

Measurement of the abundances of the four aforementioned elements in a young supernova remnant
would provide 
different estimates of the initial neutron excess, related to the progenitor metallicity, which
would allow a 
cross-check of the results. In restricted cases, it would  allow roughly estimating the expansion
timescale 
of the explosion:
\begin{itemize}
 \item Tracers of the initial neutron excess.
 \begin{enumerate}
  \item Mn/Cr is an excellent tracer of $\eta$, whose best fit is given by Eq.~\ref{eqfitmncr}.
  \item V/Ti is a good tracer of $\eta$, although with high dispersion at the lowest values of the
neutron excess 
        explored. The best fit is given by Eq.~\ref{eqfitvti}.
  \item Mn/Ti is an excellent tracer of $\eta$, whose best fit is given by Eq.~\ref{eqfitmnti}.
 \end{enumerate}
 \item Tracer of the expansion timescale.
 \begin{enumerate}
  \item V/Mn can be used to obtain a rough estimate of the expansion timescale only if the neutron
excess has 
        been measured previously and if the corresponding progenitor metallicity is lower than
$\sim0.1 Z_{\sun}$. In this case, the
        best fit is given by Eq.~\ref{eq26}.
 \end{enumerate}
 \item Control variable.
 \begin{enumerate}
  \item Ti/Cr barely depends on any parameter of Si-b, so it can be used to check that the
present 
        predictions are correct or to check that the reverse shock is inside the incomplete Si-b
zone. The
        value of Ti/Cr we found is in the range $0.035 - 0.06$.
 \end{enumerate}
\end{itemize}

The mass of each one of these elements in the ejecta conditions their detectability. Although the 
absolute mass may vary from event to event, a typical mass of chromium is $\sim 5 -
7\times10^{-3}$~M$_{\sun}$, 
and the mass of the other three elements can be deduced from Figs.~\ref{fig9} and
\ref{fig11}-\ref{fig15}. 
Titanium mass is one to two orders of magnitude less than chromium mass, whereas the mass of
manganese oscillates 
between two orders of magnitude less and the same order as the chromium mass, depending on the
neutron excess. 
Vanadium is the least prolific element, with masses two to four orders of magnitude less than the
chromium mass.

Chromium and manganese have already been detected with {\sl Suzaku} in a number of SNIa remnants 
\citep{yam10,yam12}. The detection of vanadium in X-rays is problematic due to its low abundance,
whereas that of 
titanium may be made difficult by superposition of the emission from non-dominant ions of lavishly
produced calcium,
i.e. those emitting at a higher energy than the bulk of calcium. The emission peaks of titanium to
manganese 
are separated well in energy, by more than $\sim300$~eV, which ensures that their features can be
resolved by the 
instrument XIS onboard {\sl Suzaku}, as well as with future missions. 
For instance, the {\sl IXO} proposal contemplated a resolution of 6 eV at 6 keV, with an effective
area of 
$0.65$~m$^2$, which is $\sim6.5$ times better than the effective area of XIS at the same energy.
Thus, one can expect that detection 
of chromium and manganese, and perhaps titanium, in supernova remnants will be common in the
future. 

A quite different question is the quantitative interpretation of X-ray observations of supernova
remnants, needed to 
obtain the desired abundance ratios. First, not all the atomic data needed to properly 
model X-ray emission from these elements are available at present. Second, it is difficult to
estimate the physical 
variables that influence X-ray spectra. Hopefully, technical advances in the near 
future will make this interpretation more reliable.

\subsection{Photospheric epoch of SNIa}

The photospheric epoch of SNIa starts shortly after the explosion, and ends approximately a month
later. It shares
a common picture with young supernova remnant, in the sense that the optical emission is
determined mainly by the position 
of a receding front, in this case the photosphere. While the
photosphere remains far enough from the center of the ejecta, the optical emission thus comes from
matter whose chemical 
composition was set by the incomplete Si-b process.
In this epoch, it is necessary to measure the abundance of manganese to obtain any
estimate of the initial neutron excess. 
\begin{itemize}
 \item Tracers of the initial neutron excess.
 \begin{enumerate}
  \item Mn/Cr again provides the best measurement of $\eta$,  because both their ratios vary
little with time, and
        it correlates strongly with the neutron excess. The best fit, which is applicable after day
five
        after the explosion, is given by Eq.~\ref{eq22}.
  \item V/Mn varies with time owing to the different contributions of radioactive isotopes of
vanadium. The best fits are given by Eq.~\ref{eq23} (day 15)
and Eq.~\ref{eq24} (from day 30 on).
  \item Ti/Mn can be fit as a single function of $\eta$, Eq.~\ref{eq25}, from day 10 on.
 \end{enumerate}
 \item Control variable.

% \begin{enumerate}
  Ti/Cr varies with time but is insensitive to any of the parameters of incomplete Si-b
(Fig.~\ref{fig12}).
% \end{enumerate}
\end{itemize}

\cite{hat99} studied the ion signatures in SNIa spectra, concluding that there are several
isotopes from  
titanium to manganese that are potentially identifiable in the photospheric epoch, namely 
\ion{Ti}{ii}, \ion{V}{ii}, \ion{Cr}{i}, \ion{Cr}{ii}, \ion{Mn}{i}, and \ion{Mn}{ii}. However,  
the only elements detected so far are titanium and chromium, both in SN2002bo and SN2004eo
\citep{ste05b,maz08}. 
%Their pressence could not be checked in SN2003du because the intersting wavelenght range was not
%covered by the spectra 
%recorded from this supernova \citep{tan10b}. 
Continued improvement in the techniques used in the recording and interpretation of SNIa spectra,
will provide better estimates of the mass of titanium, chromium, and manganese in future
observations
\footnote{
It needs to be stressed that the proper measurement of the masses of these elements
depends on accurate knowledge and modeling of their ionization state, which usually plays a
dominant role in the determination of their line strengths (S. Blondin, private communication).
}.
Obviously, the next SNIa exploding in the Milky Way will provide the opportunity to measure its
chemical composition with unprecedented detail. 

\subsection{Final remarks}

The use of the proposed abundance ratios to estimate SNIa properties is robust, in the sense that
\begin{enumerate}
 \item The profile of each one of the abundance ratios is homogeneous throughout the region
affected by 
       incomplete Si-b. As a result, it does not depend on the precise position of the reverse
shock (X-ray epoch) or of the
photosphere 
       (optical epoch).
 \item A control variable is defined, which is the ratio Ti/Cr that, if measured, would allow
checking
that the emitting
       region is within the incomplete Si-b layers.
 \item In SNIa models, titanium and chromium are synthesized mainly by the incomplete Si-b
process. 
       Manganese can have an additional contribution from NSE regions, whose relevance depends on
both the central 
       density of the progenitor WD at thermal runaway and the initial development of the
combustion wave.
       Vanadium can only take contributions from a slightly wider range of peak temperatures than
explored here.  
\end{enumerate}

It is important  that the empirical correlations between abundance ratios and initial neutron excess, found 
as a result of the integration of the nuclear evolutionary equations, are supported by a theoretical 
analysis based on the properties of QSE groups in Si-b. This theoretical analysis allows us
to understand the 
conditions under which one can expect the empirical correlations to apply. It has also been shown that the 
relatively low entropy of incomplete Si-b in SNIa is a necessary condition for the tight correlation between the
abundance ratio  Mn/Cr and the initial neutron excess. This correlation therefore cannot be
extrapolated to
incomplete Si-b matter in core-collapse supernovae.

\bibliographystyle{aa}
%\bibliography{../ebg}

\end{document}